\documentclass[aps,prd,twocolumn,superscriptaddress,floatfix,nofootinbib]{revtex4}
\usepackage{longtable}
\usepackage{txfonts}
\usepackage{bm}
\usepackage{color}
\usepackage{graphicx}
\newcommand{\be}{\begin{equation}}
\newcommand{\ee}{\end{equation}}
\newcommand{\een}{\end{subequations}}
\newcommand{\ben}{\begin{subequations}}
\newcommand{\beq}{\begin{eqalignno}}
\newcommand{\eeq}{\end{eqalignno}}

\newcommand{\lsim}{\mathrel{\mathop{\kern 0pt \rlap
      {\raise.2ex\hbox{$<$}}}\lower.9ex\hbox{\kern-.190em $ \sim$}}}
\newcommand{\gsim}{\mathrel{\mathop{\kern 0pt
      \rlap{\raise.2ex\hbox{$>$}}}\lower.9ex\hbox{\kern-.190em $\sim$}}}

\newcommand{\erf}{\mbox{erf}}

\DeclareMathAlphabet{\mathbbold}{U}{bbold}{m}{n}   
\newcommand{\CO}{\mathcal{O}}
\def\nn{\nonumber}

\begin{document}	

\title{Probing DAMA/LIBRA in the full parameter space of WIMP
  effective models of inelastic scattering}
\author{Sunghyun Kang}
\email{francis735@naver.com}
\affiliation{Department of Physics, Sogang University, 
Seoul, Korea, 121-742}

\author{S. Scopel}
\email{scopel@sogang.ac.kr}
\affiliation{Department of Physics, Sogang University, 
Seoul, Korea, 121-742}
\author{Gaurav Tomar}
\email{tomar@sogang.ac.kr}
\affiliation{Department of Physics, Sogang University, 
Seoul, Korea, 121-742}
\date{\today}

\begin{abstract}
We discuss the compatibility of the combined annual modulation effect
measured by DAMA/LIBRA--phase1 and DAMA/LIBRA--phase2 with an
explanation in terms of inelastic scattering events induced by the
most general Galilean-invariant effective contact interaction of a
Weakly Interacting Massive Particle (WIMP) dark matter particle of
spin 0, 1/2 or 1. We take into account all the possible interferences
among operators by studying the intersections among the ellipsoidal
surfaces of constant signal of DAMA and other experiments in the space
of the coupling constants of the effective theory. In our analysis we
assume a standard Maxwellian velocity distribution in the Galaxy.  We
find that, compared to the elastic case, inelastic scattering
partially relieves but does not eliminate the existing tension between
the DAMA effect and the constraints from the null results of other
experiments. Such tension is very large in all the parameter space
with the exception of a small region for WIMP mass $m_{\chi}\simeq$ 10
GeV and mass splitting $\delta\gsim$20 keV, where it is partially, but
not completely relieved. In such region the bounds from fluorine
targets are evaded in a kinematic way because the minimal WIMP
incoming speed required to trigger upscatters off fluorine exceeds the
maximal WIMP velocity in the Galaxy, or is very close to it. As a
consequence, we also find that the residual tension between DAMA and
other results is considerably more sensitive on the astrophysical
parameters compared to the elastic case. We find that the
configurations with the smallest tension can produce enough yearly
modulation in some of the DAMA bins in compliance with the constraints
from other experiments, but the ensuing shape of the modulation
spectrum is too steep compared to the measured one. For such
configurations the recent COSINE--100 bound is evaded in a natural way
due to their large expected modulation fractions.
\end{abstract}


\maketitle


\maketitle

\section{Introduction}
\label{sec:introduction}

For more than 15 years the DAMA
collaboration~\cite{dama_2008,dama_2010,dama_2013,dama_2018} has been
measuring a yearly modulation effect in a large--mass low--background
sodium iodide target compatible to the signal of the Dark Matter (DM)
particles that are believed to make up 27\% of the total mass density
of the Universe~\cite{planck} and more than 90\% of the halo of our
Galaxy.  Indeed, Weakly Interacting Massive Particles (WIMPs), which
provide one of the most popular DM explanations, are expected to have
feeble interactions with nuclear targets in a terrestrial detector
with a scattering rate that presents a modulation with a period of one
year due to the Earth revolution around the Sun~\cite{Drukier:1986tm}.
In particular, with the release of the latest DAMA/LIBRA-phase2
data~\cite{dama_2018} the statistical significance of DAMA effect has
reached almost $12\sigma$. However, in the most popular WIMP scenarios
used to explain the DAMA signal as due to WIMPs, the DAMA modulation
appears incompatible with the results from many other DM experiments
that have failed to observe any signal so far. Nevertheless, until
recently none of the experiments ruling out the DAMA effect used the
same target nuclei as DAMA/LIBRA, so that such incompatibility relied
on both Particle--Physics and Astrophysics assumptions. Such model
dependence has been shown to persist~\cite{psidm_2018} also after the
bound from the COSINE--100 collaboration~\cite{cosine_nature}, that
has recently published an exclusion plot for a standard elastic,
spin--independent isoscalar WIMP nucleus interaction and a WIMP
Maxwellian velocity distribution that for the first time rules out the
DAMA effect at low WIMP masses using 106 kg of $NaI$, the same target
of DAMA.  Given the strong statistical significance of the DAMA/LIBRA
signal, and the scientific implications, this prompted the need to
extend the class of WIMP models. Indeed, several scenarios have been
introduced trying to reconcile the DAMA effect with other null
results~\cite{kurylov_kamionkowski_2004,
  gelmini_gondolo_2004,mirror_dm_2008,wimpless_2008,susy_2008,composite_2008,
  inelastic_2005,gelmini_gondolo_2009,resonant_dm_2009,
  freese_gondolo_2009,leptophilic_dm_2009,luminous_dm_2010,
  spin_inelastic_2010,delnobile_gelmini_2015,
  anapole_2018,psidm_2018}. A more systematic approach is to compare
DAMA and other null results exploiting the non--relativistic (NR)
nature of the WIMP--scattering process, that allows to express the
interaction in terms of the more general effective Hamiltonian allowed
by Galilean invariance~\cite{haxton1,haxton2}, of the form:

\begin{eqnarray}
{\bf\mathcal{H}}({\bf{r}})&=& \sum_{\tau=0,1} \sum_{j=1}^{15} c_j^{\tau} \mathcal{O}_{j}({\bf{r}}) \, t^{\tau} ,
\label{eq:H}
\end{eqnarray}
\noindent
where $t^{0}= \mathbbold{1}$ is the identity in isospin space, $t^{1}=
\tau_3$ is the third Pauli matrix, and ${\mathbf{r}}$ denotes the dark
matter-nucleon relative distance and the operators $\mathcal{O}_{j}$
depend on the exchanged momentum $\vec{q}$, the WIMP incoming velocity
$\vec{v}$, the WIMP spin $\vec{S}_{\chi}$ and nuclear spin
$\vec{S}_{N}$.  In Eq.~(\ref{eq:H}) the isoscalar and isovector
(dimension -2) coupling constants $c^0_j$ and $c^{1}_j$, are related
to those to protons and neutrons $c^{p}_j$ and $c^{n}_j$ by
$c^{p}_j=(c^{0}_j+c^{1}_j)$ and $c^{n}_j=(c^{0}_j-c^{1}_j)$.
Truncating the effective theory expansion to operators at most
quadratic in $q\equiv |\vec{q}|$ and $v\equiv |\vec{v}|$ the ensuing
Hamiltonian contains 8 independent couplings $c_j^{\tau}$ for a scalar
WIMP, 28 independent couplings for a spin--1/2
WIMP~\cite{haxton1,haxton2} and 20 couplings for
spin--1~\cite{krauss_spin_1}. The ultimate assessment of the
compatibility of DAMA with other constraints requires a full
exploration of such large parameter space, where the relative
sensitivities of different nuclear targets to DM scattering events may
vary by orders of magnitude. While, due to the large dimensionality,
its direct scanning appears to be challenging this has been achieved
by using matricial techniques~\cite{Catena_dama}, exploiting the fact
that in terms of the couplings vector ${\bf c}=\left(c_1^{(0)},
c_1^{(1)},..., c_{n}^{(0)}, c_{n}^{(1)}\right)^T$ for all direct
detection experiments the expected event rate can be written in the
form:

\begin{equation}
\textrm{event rate} \, \propto \, {\bf c}^T \varmathbb{X}  \;{\bf c} \,.
\label{eq:quadratic_dependence}
\end{equation}
with $\varmathbb{X}$ is a real symmetric $2 n \times 2 n$ matrix,
which encodes all the information about nuclear responses, the dark
matter velocity distribution, experimental efficiencies, etc., but
which is \emph{independent} of the underlying particle physics model
(for a given dark matter mass). Due to this factorization in the
effective field theory parameter space the surfaces of constant signal
in different detectors are ellipsoids, and, as discussed in
Refs.~\cite{Catena_dama,Catena_halo_indep_eft}, the determination of
their geometrical intersections allows to efficiently compare the
results of various direct detection experiments in the
high-dimensional parameter space of the non-relativistic effective
theory, without making any a priori assumptions regarding the relative
size of the various Wilson coefficients $c_k^\tau$. In this way for a
standard halo model the DAMA/LIBRA--phase1
result~\cite{dama_2008,dama_2010,dama_2013} was shown to be
incompatible to the constraints from other experiments in the case of
elastic interactions of a WIMP particle with spin $\le$ 1/2 and the
interaction Hamiltonian of Eq.(~\ref{eq:H}) with arbitrary couplings
combinations~\cite{Catena_dama}.

As pointed out by the authors, the analysis in~\cite{Catena_dama} did
not cover several alternative scenarios. One of them is inelastic
scattering. Indeed, one of the very few scenarios that reconcile DAMA
with the constraints of other experiments is proton--philic
Spin--dependent Inelastic Dark Matter
(pSIDM)~\cite{psidm_2015,psidm_2017,psidm_2018}. In such model the
WIMP particle interacts with nuclear targets through a spin--dependent
coupling that is suppressed on neutrons, in order to comply to
constraints using neutron--odd scattering targets (germanium and
xenon).  Moreover inelastic scattering
(IDM)~\cite{inelastic,inelastic_2009,inelastic_2009_velocity_dist}
reconciles the above scenario to fluorine detectors. In IDM a DM
particle $\chi_1$ of mass $m_{\chi_1}=m_{\chi}$ interacts with atomic
nuclei exclusively by up--scattering to a second heavier state
$\chi_2$ with mass $m_{\chi_2}=m_{\chi}+\delta$. A peculiar feature of
IDM is that there is a minimal WIMP incoming speed in the lab frame
matching the kinematic threshold for inelastic upscatters and given
by:

\begin{equation}
v_{min}^{*}=\sqrt{\frac{2\delta}{\mu_{\chi N}}},
\label{eq:vstar}
\end{equation}

\noindent with $\mu_{\chi N}$ the WIMP--nucleus reduced
mass. This quantity corresponds to the lower bound of the minimal
velocity $v_{min}$ (also defined in the lab frame) required to deposit
a given recoil energy $E_R$ in the detector:

\begin{equation}
v_{min}=\frac{1}{\sqrt{2 m_N E_R}}\left | \frac{m_NE_R}{\mu_{\chi N}}+\delta \right |,
\label{eq:vmin}
\end{equation}

\noindent with $m_N$ the nuclear mass.

In particular, indicating with $v_{min}^{*Na}$ and $v_{min}^{*F}$ the
values of $v_{min}^*$ for sodium and fluorine, and with $v_{esc}$ the
WIMP escape velocity, in Refs.~\cite{psidm_2015,psidm_2017,psidm_2018}
constraints from WIMP--fluorine scattering events in droplet detectors
and bubble chambers were shown to be evaded when the WIMP mass
$m_{\chi}$ and the mass gap $\delta$ are chosen in such a way that the
hierarchy:

\begin{equation}
v_{min}^{*Na}<v_{esc}^{lab}<v_{min}^{*F},
\label{eq:hierarchy}
\end{equation}

\noindent is achieved. In fact, in such case WIMP scatterings off
fluorine turn kinematically forbidden while those off sodium can still
serve as an explanation to the DAMA effect. So the pSIDM mechanism
rests on the trivial observation that the velocity $v_{min}^*$ for
fluorine is larger than that for sodium.

In the present paper we wish to apply the technique introduced in
Ref.~\cite{Catena_dama} to extend the analyses
of~\cite{psidm_2015,psidm_2017,psidm_2018} to the general interaction
Hamiltonian of Eq.~(\ref{eq:H}), or, conversely, we wish to extend the
analysis of \cite{Catena_dama} to the case of inelastic scattering,
updating it to the present experimental situation and including the
DAMA/LIBRA--phase2 data release and the state of the art of all the
constraints from other experiments.

The two most important improvements of DAMA/LIBRA-phase2 compared to
the previous phases are that now the exposure has almost doubled and
the energy threshold has been lowered from 2 keV electron-equivalent
(keVee) to 1 keVee. In particular this latter feature introduces an
important difference between the present analysis and that of
Ref.~\cite{Catena_dama}. In the latter, for kinematic reasons and
irrespective of the effective interaction the DAMA/LIBRA-phase1 data
were only sensitive to scattering events off Sodium for a WIMP mass
$m_{\chi}\lsim$ 20 GeV, implying that in such range of mass the
scaling law induced by the effective Hamiltonian~(\ref{eq:H}) only
entered in the comparison of the scattering rate off Sodium and that
off the targets of other experiments. However, due to the lower
threshold, now for $m_{\chi}\lsim$ 20 GeV DAMA/LIBRA--phase2 is
sensitive to both the target nuclei, with WIMP--Iodine scattering
events contributing to the expected rate in the new low--energy range
below 2 keVee and Sodium at higher energy. This implies that the scaling
law among different targets is now also relevant in explaining the
energetic spectrum of the modulation amplitudes measured by DAMA
alone. Indeed, due to this reason for a standard Maxellian velocity
distribution the the goodness--of--fit of a WIMP explanation of the
DAMA/LIBRA-phase2 data has already been showed to worsen compared to
DAMA/LIBRA-phase1 for a standard Spin-Independent interaction
(SI)~\cite{dama_phase2_first_analysis,freese_2018}, requiring to tune
the ratio between the WIMP--proton and the WIMP--neutron couplings in
order to suppress WIMP--Iodine scattering events below 2 keVee.  As
shown in~\cite{dama_2018_sogang}, with the exception of
$\mathcal{O}_{1}$ and $\mathcal{O}_{4}$, which in the notation
of~\cite{haxton1,haxton2} correspond respectively to the standard SI
or Spin--Dependent (SD) interactions, this problem is not present for
all the other operators $\mathcal{O}_{j}$ of the
Hamiltonian~(Eq.~\ref{eq:H}). This represents an additional motivation to
update the result of Ref.~\cite{Catena_dama} to the
DAMA/LIBRA-phase2 data. Finally, the recent COSINE--100 bound is
potentially relevant to our analysis because any probe of the DAMA
effect using $NaI$ is expected to lead to conclusions independent on
the WIMP--nucleus cross section scaling law and so on the particular
choice of the couplings of the Hamiltonian~(Eq.~\ref{eq:H}).  However, as
we will show, an important dependence on the Hamiltonian~(Eq.~\ref{eq:H})
is still present when comparing COSINE-100 and DAMA.  This is due to
the fact that, although an initial modulation analysis of COSINE-100
with two--year data is forthcoming and an additional low--threshold
analysis is also actively under development, COSINE-100 needs to
collect several years of data~\cite{Thompson:2017yvq} in order to
reach the sensitivity required to probe the DAMA signal, and until
then COSINE--100 will only exploit the average count rate.
So the results of the two experiments are presently based on two
different observables, the yearly modulation expected from the rotation
of the Earth around the Sun and the time--averaged
rate, and their relative size does depend on the specific model of
WIMP--nucleus interaction besides a standard SI or SD interaction with
nuclei.

The paper is organized as follows: in Section~\ref{sec:model} we
summarize how we calculate WIMP direct detection rates in NR effective
theory; in Section~\ref{sec:maximal}
we outline the geometrical method of Ref.~\cite{Catena_dama} that we use
to study the intersections among the ellipsoidal surfaces of constant
signal of DAMA and other experiments in the space of the coupling
constants of the effective theory; our quantitative analysis is
contained in Section~\ref{sec:analysis}. We devote
Section~\ref{sec:conclusions} to our conclusions. For completeness we
summarize the response functions of a WIMP of spin $\le$1 in
Appendix~\ref{app:wimp_eft} and we provide the details of our
treatment of experimental constraints in Appendix~\ref{app:exp}.

\section{WIMP inelastic scattering in non-relativistic effective models}
\label{sec:model}

In the present Section we briefly summarize the ingredients that we
use to calculate for each experiment and for each energy bin used in
our analysis the matrix $\varmathbb{X}$ introduced in
Eq.~(\ref{eq:quadratic_dependence}), needed to evaluate the expected
rate to compare to the experimental data.

The full list of operators $\mathcal{O}_{j}$ entering the Hamiltonian
of Eq.~\ref{eq:H} for the nuclear scattering process of a WIMP
particle of spin $J{\chi}\le$ 1 is given by:

\begin{eqnarray}
  \CO_1 &=& 1_\chi 1_N; \; \CO_2 = (v^\perp)^2; \;\;\;\;\nn\\
  \CO_3 &=& i \vec{S}_N \cdot ({\vec{q} \over m_N} \times \vec{v}^\perp);\; \CO_4 = \vec{S}_\chi \cdot \vec{S}_N;\; \nn\\
  \CO_5 &=& i \vec{S}_\chi \cdot ({\vec{q} \over m_N} \times \vec{v}^\perp);\; \CO_6=
  (\vec{S}_\chi \cdot {\vec{q} \over m_N}) (\vec{S}_N \cdot {\vec{q} \over m_N}) \nn \\
  \CO_7 &=& \vec{S}_N \cdot \vec{v}^\perp;\;\;\CO_8 = \vec{S}_\chi \cdot \vec{v}^\perp;\;\nn\\
  \CO_9 &=& i \vec{S}_\chi \cdot (\vec{S}_N \times {\vec{q} \over m_N});\; \CO_{10} = i \vec{S}_N \cdot {\vec{q} \over m_N};\;\nn\\
  \CO_{11} &=& i \vec{S}_\chi \cdot {\vec{q} \over m_N};\;\CO_{12} = \vec{S}_\chi \cdot (\vec{S}_N \times \vec{v}^\perp) \nn\\
  \CO_{13} &=&i (\vec{S}_\chi \cdot \vec{v}^\perp  ) (  \vec{S}_N \cdot {\vec{q} \over m_N});\;\CO_{14} = i ( \vec{S}_\chi \cdot {\vec{q} \over m_N})(  \vec{S}_N \cdot \vec{v}^\perp )  \nn\\
  \CO_{15} &=& - ( \vec{S}_\chi \cdot {\vec{q} \over m_N}) ((\vec{S}_N \times \vec{v}^\perp) \cdot {\vec{q} \over m_N});\;\nn\\
  \CO_{16} &=& - ((\vec{S}_\chi \times \vec{v}^\perp)\cdot {\vec{q} \over m_N}) (\vec{S}_N \cdot {\vec{q} \over m_N});\;\nn\\
  \CO_{17} &=& i {\vec{q} \over m_N} \cdot \mathcal{S} \cdot \vec{v}^\perp;\;
  \CO_{18} = i {\vec{q} \over m_N} \cdot \mathcal{S} \cdot \vec{S}_N,
\label{eq:ops}
\end{eqnarray}

\noindent In the equation above $1_{\chi N}$ is the identity operator,
$\vec{q}$ is the transferred momentum, $\vec{S}_{\chi}$ and
$\vec{S}_{N}$ are the WIMP and nucleon spins, respectively, while
$\mathcal{S}=\frac{1}{2}(\epsilon^{\dagger}_i\epsilon_j+\epsilon^{\dagger}_j\epsilon_i)$
is the symmetric combination of polarization vectors in the case of a
spin--1 DM particle and $\vec{v}^\perp = \vec{v} +
\frac{\vec{q}}{2\mu_{\chi {\cal N}}}$ (with $\mu_{\chi {\cal N}}$ the
WIMP--nucleon reduced mass) is the relative transverse velocity
operator satisfying $\vec{v}^{\perp}\cdot \vec{q}=0$. For a nuclear
target $T$ the quantity $(v^{\perp}_T)^2 \equiv |\vec{v}^{\perp}_T|^2$
can also be written as~\cite{eft_inelastic}:

\begin{equation}
(v^{\perp}_T)^2=v^2_T-v_{min}^2.
\label{eq:v_perp}
\end{equation}

\noindent where $v_{min}$ is given by Eq.(\ref{eq:vmin}).

Operator ${\cal O}_2$ is of higher order in $v$ compared to all the
others, implying a cross section suppression of order ${\cal
  O}(v/c)^4)\simeq 10^{-12}$ for the non--relativistic WIMPs in the
halo of our Galaxy. Moreover it cannot be obtained from the
leading-order non relativistic reduction of a manifestly relativistic
operator \cite{haxton1}.  So, following Refs.\cite{haxton1,haxton2},
we will not include it in our analysis.  Moreover, operator $\CO_{16}$
is a linear combination of other operators, so can be omitted. This
implies a maximal number of 16 operators in the effective Hamiltonian
in Eq.~(\ref{eq:H}), namely 4 operators for a spin--0 DM particle, 14
operators for spin 1/2 and 10 operators for spin 1.

To reduce the parameter space of the effective interaction of
Eq.~(\ref{eq:H}) in the following we will make a few simplifying
assumptions. First, we will only consider the case $\delta>$0,
i.e. upscatters of a light state to a heavier one; then we will assume
a contact effective interaction between the WIMP and the nucleus,
i.e., we will assume the coefficients $c_j^{\tau}$ as independent on
the transferred momentum $q$ and neglect propagator effects. Moreover,
we will consider real $c_j^{\tau}$'s, although in general for an
inelastic process they can be complex~\cite{eft_inelastic}. Finally,
we will not consider the possibility of inelastic scattering among
states of different spins~\cite{eft_inelastic}.

The expected rate in a given visible energy bin $E_1^{\prime}\le
E^{\prime}\le E_2^{\prime}$ of a direct detection experiment is given
by:

\begin{eqnarray}
R_{[E_1^{\prime},E_2^{\prime}]}&=&M\mbox{T}\int_{E_1^{\prime}}^{E_2^{\prime}}\frac{dR}{d
  E^{\prime}}\, dE^{\prime}, \label{eq:start}\\
 \frac{dR}{d E^{\prime}}&=&\sum_T \int_0^{\infty} \frac{dR_{\chi T}}{dE_{ee}}{\cal
   G}_T(E^{\prime},E_{ee})\epsilon(E^{\prime})\label{eq:start2}\,d E_{ee}, \\
E_{ee}&=&q(E_R) E_R \label{eq:start3},
\end{eqnarray}

\noindent with $\epsilon(E^{\prime})\le 1$ the experimental
efficiency/acceptance. In the equations above $E_R$ is the recoil
energy deposited in the scattering process (indicated in keVnr), while
$E_{ee}$ (indicated in keVee) is the fraction of $E_R$ that goes into
the experimentally detected process (ionization, scintillation, heat)
and $q(E_R)$ is the quenching factor, ${\cal
  G}_T(E^{\prime},E_{ee}=q(E_R)E_R)$ is the probability that the
visible energy $E^{\prime}$ is detected when a WIMP has scattered off
an isotope $T$ in the detector target with recoil energy $E_R$, $M$ is
the fiducial mass of the detector and T the live--time of the data
taking. For a given recoil energy imparted to the target the
differential rate for the WIMP--nucleus scattering process is given
by:

\be
\frac{d R_{\chi T}}{d E_R}(t)=\sum_T N_T\frac{\rho_{\mbox{\tiny WIMP}}}{m_{\chi}}\int_{v_{min}}d^3 v_T f(\vec{v}_T,t) v_T \frac{d\sigma_T}{d E_R},
\label{eq:dr_de}
\ee

\noindent where $\rho_{\mbox{\tiny WIMP}}$ is the local WIMP mass density in the
neighborhood of the Sun, $N_T$ the number of the nuclear targets of
species $T$ in the detector (the sum over $T$ applies in the case of
more than one nuclear isotope), while

\be
\frac{d\sigma_T}{d E_R}=\frac{2 m_T}{4\pi v_T^2}\left [ \frac{1}{2 j_{\chi}+1} \frac{1}{2 j_{T}+1}|\mathcal{M}_T|^2 \right ],
\label{eq:dsigma_de}
\ee

\noindent with $m_T$ the nuclear target mass and, assuming that the
nuclear interaction is the sum of the interactions of the WIMPs with
the individual nucleons in the nucleus:

\begin{eqnarray}
&&  \frac{1}{2 j_{\chi}+1} \frac{1}{2 j_{T}+1}|\mathcal{M}_T|^2=\nonumber\\
&&  \frac{4\pi}{2 j_{T}+1} \sum_{\tau=0,1}\sum_{\tau^{\prime}=0,1}\sum_{k} R_k^{\tau\tau^{\prime}}\left [c^{\tau}_j,(v^{\perp}_T)^2,\frac{q^2}{m_N^2}\right ] W_{T k}^{\tau\tau^{\prime}}(y).
\label{eq:squared_amplitude}
\end{eqnarray}

\noindent In the above expression $j_{\chi}$ and $j_{T}$ are the WIMP
and the target nucleus spins, respectively, $q=|\vec{q}|$ while the
$R_k^{\tau\tau^{\prime}}$'s are WIMP response functions (that we
report for completeness in Eq.(\ref{eq:wimp_response_functions}))
which depend on the couplings $c^{\tau}_j$ as well as the transferred
momentum $\vec{q}$ and $(v^{\perp}_T)^2$. In equation
(\ref{eq:squared_amplitude}) the $W^{\tau\tau^{\prime}}_{T k}(y)$'s
are nuclear response functions and the index $k$ represents different
effective nuclear operators, which, crucially, under the assumption
that the nuclear ground state is an approximate eigenstate of $P$ and
$CP$, can be at most eight: following the notation in
\cite{haxton1,haxton2}, $k$=$M$, $\Phi^{\prime\prime}$,
$\Phi^{\prime\prime}M$, $\tilde{\Phi}^{\prime}$,
$\Sigma^{\prime\prime}$, $\Sigma^{\prime}$,
$\Delta$, $\Delta\Sigma^{\prime}$. The $W^{\tau\tau^{\prime}}_{T
  k}(y)$'s are function of $y\equiv (qb/2)^2$, where $b$ is the size
of the nucleus. For the target nuclei $T$ used in most direct
detection experiments the functions $W^{\tau\tau^{\prime}}_{T k}(y)$,
calculated using nuclear shell models, have been provided in
Refs.~\cite{haxton2,catena} under the assumption that the dark matter
particle couples to the nucleus through local one--body interactions
with the nucleons. In our analysis we do not include two--body
effects~\cite{two_body_1,two_body_2} which are only available for a
few isotopes and can be important when the one--body contribution is
suppressed. 
Finally, $f(\vec{v}_T)$ is the WIMP velocity distribution, for which
we assume a standard isotropic Maxwellian at rest in the Galactic rest
frame truncated at the escape velocity $u_{esc}$, and boosted to the
Lab frame by the velocity of the Earth. So for the former we assume:

\begin{eqnarray}
  f(\vec{v}_T,t)&=&N\left(\frac{3}{ 2\pi v_{rms}^2}\right )^{3/2}
  e^{-\frac{3|\vec{v}_T+\vec{v}_E|^2}{2 v_{rms}^2}}\Theta(u_{esc}-|\vec{v}_T+\vec{v}_E(t)|),\\
  N&=& \left [ \erf(z)-\frac{2}{\sqrt{\pi}}z e^{-z^2}\right ]^{-1},  
  \label{eq:maxwellian}
  \end{eqnarray}

\noindent with $z=3 u_{esc}^2/(2 v_{rms}^2)$. In the isothermal sphere
model hydrothermal equilibrium between the WIMP gas pressure and
gravity is assumed, leading to $v_{rms}$=$\sqrt{3/2}v_0$ with $v_0$
the galactic rotational velocity.

With the exception of DAMA, all the experiments included in our
analysis are sensitive to the time average of the expected rate for
which $<v_E>$=$v_{\odot}$ and $v_{\odot}$=$v_0$+12 km/sec (accounting
for a peculiar component of the solar system with respect to the
galactic rotation).  In the case of DAMA, the yearly modulation effect
is due to the time dependence of the Earth's speed with respect to the
Galactic frame, given by:

\begin{equation}
|\vec{v}_E(t)|=v_{\odot}+v_{orb}\cos\gamma \cos\left [\frac{2\pi}{T_0}(t-t_0)
  \right ],
  \end{equation}

\noindent where $\cos\gamma\simeq$0.49 accounts for the inclination of
the ecliptic plane with respect to the Galactic plane, $T_0$=1 year
and $v_{orb}$=2$\pi r_{\oplus}/(T_0)\simeq$ 29 km/sec ($r_{\oplus}$=1
AU neglecting the small eccentricity of the Earth's orbit around the
Sun).

In our analysis for the two parameters $v_0$ and $u_{esc}$ we take
$v_0$=220 km/sec \cite{v0_koposov} and $u_{esc}$=550 km/sec
\cite{vesc_2014} as reference values, although in
Section~\ref{sec:analysis} we will also discuss the dependence of the
results when the same parameters are varied in the ranges
$v_0$=(220$\pm$ 20) km/s~\cite{v0_koposov} and $u_{esc}$=(550$\pm$ 30)
km/s~\cite{vesc_2014}. Our reference choice of parameters corresponds
  to the WIMP escape velocity in the lab rest frame
  $v_{esc}^{lab}\simeq$ 782 km/s.  To make contact with other
  analyses, for the dark matter density in the neighborhood of the Sun
  we use $\rho_{\mbox{\tiny WIMP}}$=0.3 GeV/cm$^3$, which is a standard value
  commonly adopted by experimental collaborations, although
  observations point to the slightly higher value $\rho_{\mbox{\tiny
      WIMP}}$=0.43 GeV/cm$^3$~\cite{rho_DM_salucci_1, rho_DM_salucci_2}. Notice
  that direct detection experiments are only sensitive to the product
  $\rho_{\mbox{\tiny WIMP}}\sigma_p$, so the results of the next
  Section can be easily rescaled with $\rho_{\mbox{\tiny WIMP}}$.

In particular, in each visible energy bin DAMA is sensitive to the
yearly modulation amplitude $S_m$, defined as the cosine transform of
$R_{[E_1^{\prime},E_2^{\prime}]}(t)$:

\begin{equation}
S_{m,[E_1^{\prime},E_2^{\prime}]}\equiv \frac{2}{T_0}\int_0^{T_0}
\cos\left[\frac{2\pi}{T_0}(t-t_0)\right]R_{[E_1^{\prime},E_2^{\prime}]}(t)dt,
\label{eq:sm}
\end{equation}  

\noindent with $T_0$=1 year and $t_0$=2$^{nd}$ June, while other
experiments put upper bounds on the time average $S_0$:

\begin{equation}
S_{0,[E_1^{\prime},E_2^{\prime}]}\equiv \frac{1}{T_0}\int_0^{T_0}
R_{[E_1^{\prime},E_2^{\prime}]}(t)dt.
\label{eq:s0}
\end{equation}

Using the ingredients listed above, for a given value of the two
parameters $m_{\chi}$ and $\delta$ both $S_0$ and $S_m$ can be
expressed as quadratic forms like Eq.~(\ref{eq:quadratic_dependence}),
i.e. for each of the experimental observable considered in our
analysis a real symmetric matrix can be obtained. Schematically, for
each energy bin $n$ and both for DAMA and for each of the other
experiments $exp$:

\begin{eqnarray}
  S^{DAMA}_{m,n}(m_{\chi},\delta)&=&{\bf c}^T \varmathbb{S}_{m,n}^{DAMA}(m_{\chi},\delta)  \;{\bf c}
\label{eq:matrices_sm}
  \\
S^{exp}_{0,n}(m_{\chi},\delta)&=&{\bf c}^T \varmathbb{S}_{0,n}^{exp}(m_{\chi},\delta)  \;{\bf c}.
\label{eq:matrices_s0}
\end{eqnarray}

\section{Maximal DAMA signals compatible to null results}
\label{sec:maximal}

Following the analysis in~\cite{Catena_dama}, in this section we will
use the property that, for a fixed value of the WIMP mass $m_{\chi}$
and of the mass splitting $\delta$, constant--rate surfaces in the
couplings vector space are ellipsoids. In particular, given the
experimental upper bound $N^{exp}_{0,n}$ for experiment $exp$ and
energy bin $n$, the condition $S^{exp}_{0,n}<N^{exp}_{0,n}$
implies that allowed configurations must lie inside the ellipsoid:
\begin{equation}
{\bf c}^T \varmathbb{A}_{n}^{exp}(m_{\chi},\delta){\bf c}\equiv{\bf c}^T \frac{\varmathbb{S}_{0,n}^{exp}(m_{\chi},\delta)}{N^{exp}_{0,n}} \;{\bf c}<1.
\end{equation}
\noindent As far as the DAMA modulation amplitudes are concerned, the
experimentally observed intervals
$[S^{DAMA,min}_{m,k},S^{DAMA,max}_{m,k}]$ in energy bins $k=1...N$
imply the additional upper bounds:
\begin{equation}
  {\bf c}^T\varmathbb{A}_{n}^{DAMA}(m_{\chi},\delta){\bf c}\equiv{\bf c}^T \frac{\varmathbb{S}_{m,n}^{DAMA}(m_{\chi},\delta)}{S^{DAMA,max}_{m,n}} \;{\bf c}<1,
\end{equation}
\noindent which add to the previous constraints. From now on we will
indicate all upper bound matrices as $\varmathbb{A}_{j}$, for $j\in
\varmathbb{E}$, with $\varmathbb{E}$ the full set of experimental
upper constraints including the upper bounds on the DAMA modulation
amplitudes, so that the following conditions must be verified:

\begin{equation}
  {\bf c}^T \varmathbb{A}_{j}(m_{\chi},\delta){\bf c}<1,\,\,\,j\in\varmathbb{E}.
  \label{eq:upper_bounds}
\end{equation}

\noindent In our analysis we will include the 8 DAMA modulation
amplitudes for 1 keVee $\leq E^{\prime}\leq$ 5 keVee~\cite{dama_2018}, and
selected energy bins from XENON1T~\cite{xenon_2018}, PICO--60 ($C_3F_8$
target) \cite{pico60,pico60_2019}, COSINE--100~\cite{cosine_nature},
COUPP~\cite{coupp}, SuperCDMS~\cite{super_cdms_2017} and
PICASSO~\cite{picasso}. The details of how we implemented the DAMA
effect and the bounds are provided in Appendix \ref{app:exp} and in
Table~\ref{table:exp_lmi}. Given the DAMA modulation amplitudes, an
explanation of the effect in terms of WIMPs implies also the lower
bounds:
\begin{equation}
  {\bf c}^T\varmathbb{B}_{n}(m_{\chi},\delta){\bf c}\equiv{\bf
    c}^T
  \frac{\varmathbb{S}_{m,n}^{DAMA}(m_{\chi},\delta)}{S^{DAMA,min}_{m,n}}
  \;{\bf c}>1.
    \label{eq:lower_bounds}
\end{equation}

\noindent Compatibility between DAMA and the other experiments is
achieved only if in the coupling constants parameter space the
intersection between the volumes outside the ellipsoids ${\bf
  c}^T\varmathbb{B}_{n}(m_{\chi},\delta){\bf c}$=1, $n=1,...N$ and the
volume inside the ellipsoids ${\bf
  c}^T\varmathbb{A}_{k}(m_{\chi},\delta){\bf c}$=1,
$k\in\varmathbb{E}$ is non--vanishing. To prove this it is sufficient
to find a set of real parameters $\xi_k\le 0$ that, for each DAMA
energy bin $n$ satisfy~\cite{s_lemma}:

\begin{eqnarray}
 && \sum_{i\in\varmathbb{E}}  \xi_i<1, \nonumber\\
  && \sum_{i\in\varmathbb{E}}  \xi_i \varmathbb{A}_k-\varmathbb{B}_n\,\,\,\mbox{is a positive matrix}.
  \label{eq:lmi}
\end{eqnarray}

\begin{figure}
\begin{center}
\includegraphics[width=0.5\textwidth]{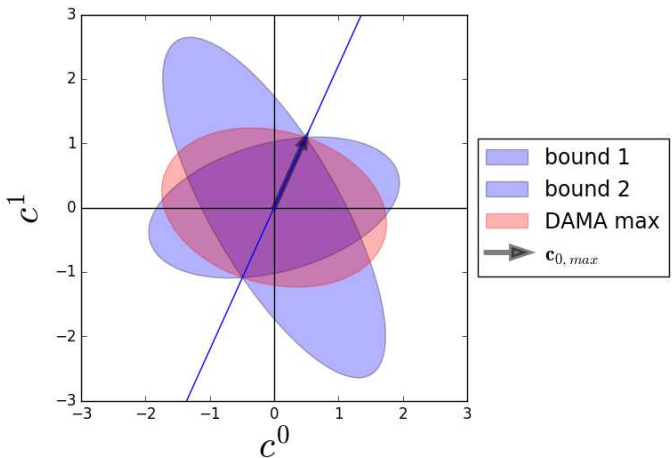}     
\end{center}
\caption{A schematic view of the intersection between the ellipse of
  maximal allowed constant modulation amplitude in one of the DAMA
  bins and the edge of the experimentally allowed volume in the
  coupling constants parameters space for the case of a single
  coupling. The solid line passing through the origin represents the
  direction singled out by the normalized eigenvector ${\bf
    \hat{c}}_{0}$ of Eq.~(\ref{eq:eigen}), while the arrow joining the
  origin to one of the two intersections represents the vector ${\bf
    c}_{0,max}$ of Eq.~(\ref{eq:c_max}) containing the set of couplings
  for the configuration that maximizes the modulation amplitude.
\label{fig:ellipses_example}}
\end{figure}

\noindent In particular, if the binary test above is verified no set
of couplings ${\bf c}$ exists for which the two conditions
(\ref{eq:upper_bounds}) and (\ref{eq:lower_bounds}) are satisfied at
the same time. Geometrically, this implies that the volume of
intersection among the ellipsoids $\varmathbb{A}_k$ is fully contained
in the DAMA ellipsoid of $\varmathbb{B}_n$. On the other hand, when
the matrix of Eq.~(\ref{eq:lmi}) is not positive--defined such set of
couplings exists. Notice that, in such case, since the
$\varmathbb{A}_k$ matrices include the upper bounds on the DAMA
modulation amplitudes, for that choice of $(m_{\chi},\delta)$ the
condition $S^{DAMA,min}_{m,n}\in
[S^{DAMA,min}_{m,n},S^{DAMA,max}_{m,n}]$ is automatically satisfied in
the energy bin $n$ in compliance to all other existing constraints,
although this is not guaranteed for the modulation amplitudes in the
other energy bins. An alternative way to show the result of the test
above is to calculate the maximal value of the modulation amplitude
allowed by present constraints, i.e. to take $S^{DAMA,max}_{m,n}$ in
$\varmathbb{B}_n$ as a free parameter and find the minimal value
$\hat{S}^{DAMA,max}_{m,n}$ for which the condition~(\ref{eq:lmi}) is
verified~\cite{Catena_dama,Catena_halo_indep_eft}.  A schematic view
of the intersection between the ${\bf c}^T
\varmathbb{S}_{m,n}^{DAMA}(m_{\chi},\delta)/\hat{S}^{DAMA,max}_{m,n}
\;{\bf c}=1$ ellipse and the edge of the experimentally allowed volume
in the coupling constants space is provided in
Fig.~\ref{fig:ellipses_example} for the case of a single coupling and
two upper bounds.  Such value can then be converted in a number of
standard deviations $n_{\sigma}$ away from the measurement (in
absolute value). The tension between DAMA and the other experiments
can then be quantified as the maximum of
$N_{\sigma}\equiv\max(n_{\sigma})$ among the DAMA energy bins
calculated in the following way: i) we fix one target energy bin $n$;
ii) we maximize the modulation signal in $n$; iii) for the
corresponding set of couplings we calculate the modulation signal also
in the other bins; iv) we take $n_\sigma$ as the maximum tension among
all the bins; v) we loop over the target bin $n$ and take the minimum
(since each target bin yields a different model).  Notice that
the procedure described above may not yield the model which better
reproduces the data in the bins where the maximal allowed modulation
exceeds the central value of the measurement.

In total, we have solved Eq.~(\ref{eq:lmi}) using 27 matrices (8+8
DAMA matrices plus 11 matrices for null results).

\section{Analysis}
\label{sec:analysis}
In this section we discuss the $N_{\sigma}$ parameter solving
Eq.~(\ref{eq:lmi}) using PICOS~\cite{picos}, a Python interface to
conic optimization, together with the CVXOPT solver~\cite{cvxopt}.

\begin{figure}
\begin{center}
\includegraphics[width=0.5\textwidth]{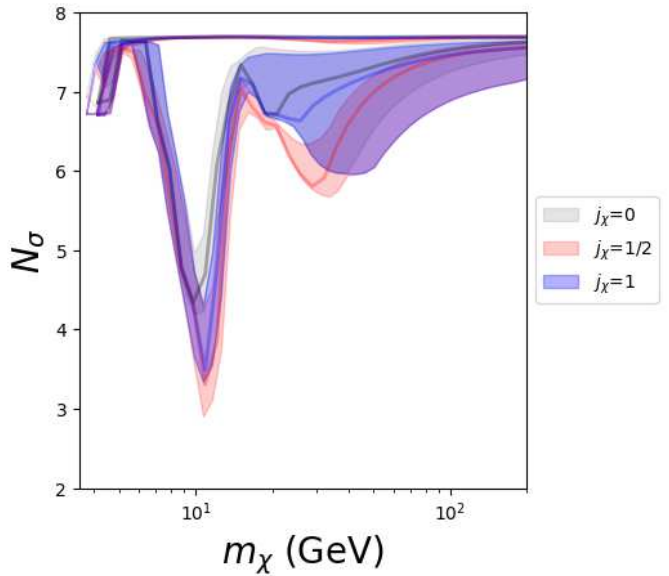}
\end{center}
\caption{Tension $N_{\sigma}$ (maximized among DAMA bins for 1
  keVee$\le E^{\prime}\le$5 keVee) between the 90\% C.L. lower bound
  of the measured modulation fractions and the maximal value of the
  same quantity allowed by 90\% C.L. upper bounds from null results.
  Solid lines show the results when the parameters $v_0$ and $u_{esc}$
  of the Maxwellian distribution of Eq.~(\ref{eq:maxwellian}) are
  fixed to the reference values $v_0$=220 km/s and $u_{esc}$=550 km/s,
  while the bands indicate the variation of $N_{\sigma}$ when $v_0$
  and $u_{esc}$ are varied in the ranges $v_0$=(220$\pm$ 20)
  km/s~\cite{v0_koposov} and $u_{esc}$=(550$\pm$ 30)
  km/s~\cite{vesc_2014}. The gray shaded regions represent
  $j_{\chi}$=0, the red bands $j_{\chi}$=1/2 and the purple ones
  represents $j_{\chi}$=1. For each value of $j_{\chi}$ the upper band
  represents the elastic case ($\delta$=0), while the lower one the
  inelastic case, when $N_{\sigma}$ is minimized in terms of
  $\delta$. at fixed $m_{\chi}$.}
\label{fig:lmi_mchi}
\end{figure}

The main results of our analysis is shown in Fig.~\ref{fig:lmi_mchi}.
In such figure, for different value of the WIMP spin $j_{\chi}$, the
upper band shows $N_{\sigma}$ as a function of $m_{\chi}$ in the
elastic case ($\delta$=0), while the lower bands represent
$N_{\sigma}$ minimized in terms of $\delta$ at fixed $m_{\chi}$.  The
bands indicate the variation of $N_{\sigma}$ when the parameters $v_0$
and $u_{esc}$ of the Maxwellian distribution of
Eq.~(\ref{eq:maxwellian}) are varied in the ranges $v_0$=(220$\pm$ 20)
km/s~\cite{v0_koposov} and $u_{esc}$=(550$\pm$ 30)
km/s~\cite{vesc_2014}, while the solid line indicates the result for
the reference values $v_0$=220 km/s and $u_{esc}$=550 km/s. The gray
shaded regions represent $j_{\chi}$=0, the red bands $j_{\chi}$=1/2
and the purple ones represent $j_{\chi}$=1. As far as the $\delta$=0
case is concerned, a DAMA explanation is excluded at more than
$\simeq$ 7 sigmas for all WIMP masses below 200 GeV. Compared to the
elastic case, inelastic scattering partially relieves this tension
with values as low as $\simeq$ 4.0$\sigma$ for $j_{\chi}$=0, $\simeq$
2.9$\sigma$ for $j_{\chi}$=1/2 and $\simeq$ 3.2$\sigma$ for
$j_{\chi}$=1.  However $N_{\sigma}$ is considerably more sensitive on
the astrophysical parameters $v_0$ and $u_{esc}$ compared to the
elastic case. From this figure one can conclude that neither the large
range of different interactions provided by the effective field theory
nor the modified kinematics due to inelasticity can eliminate
completely the tension between DAMA and experimental constraints.
\begin{figure}
\begin{center}
\includegraphics[width=0.5\textwidth]{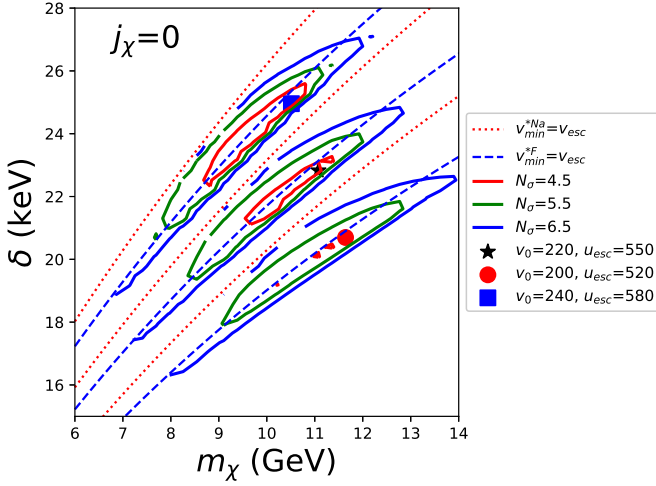}
\end{center}
\caption{Contour plots of $N_{\sigma}$ in the the $m_{\chi}$--$\delta$
  plane for $j_{\chi}$=0, from $N_{\sigma}$=4.5 to $N_{\sigma}$=6.5
  from inside out. The points of minimal $N_{\sigma}$ are represented
  by a star, a circle and a square for
  ($v_0$,$u_{esc}$)=(220,550),(200,520),(240,580) km/s, respectively,
  and surrounded by the corresponding contour plots of $N_{\sigma}$.
  For each ($v_0$,$u_{esc}$) combination the dotted (red) line
  represents $v^{*Na}_{min}$=$v_{esc}^{lab}$, while the (blue) short
  dashes show $v^{*F}_{min}$=$v_{esc}^{lab}$. }
\label{fig:mchi_delta_spin_0}
\end{figure}

\begin{figure}
\begin{center}
\includegraphics[width=0.5\textwidth]{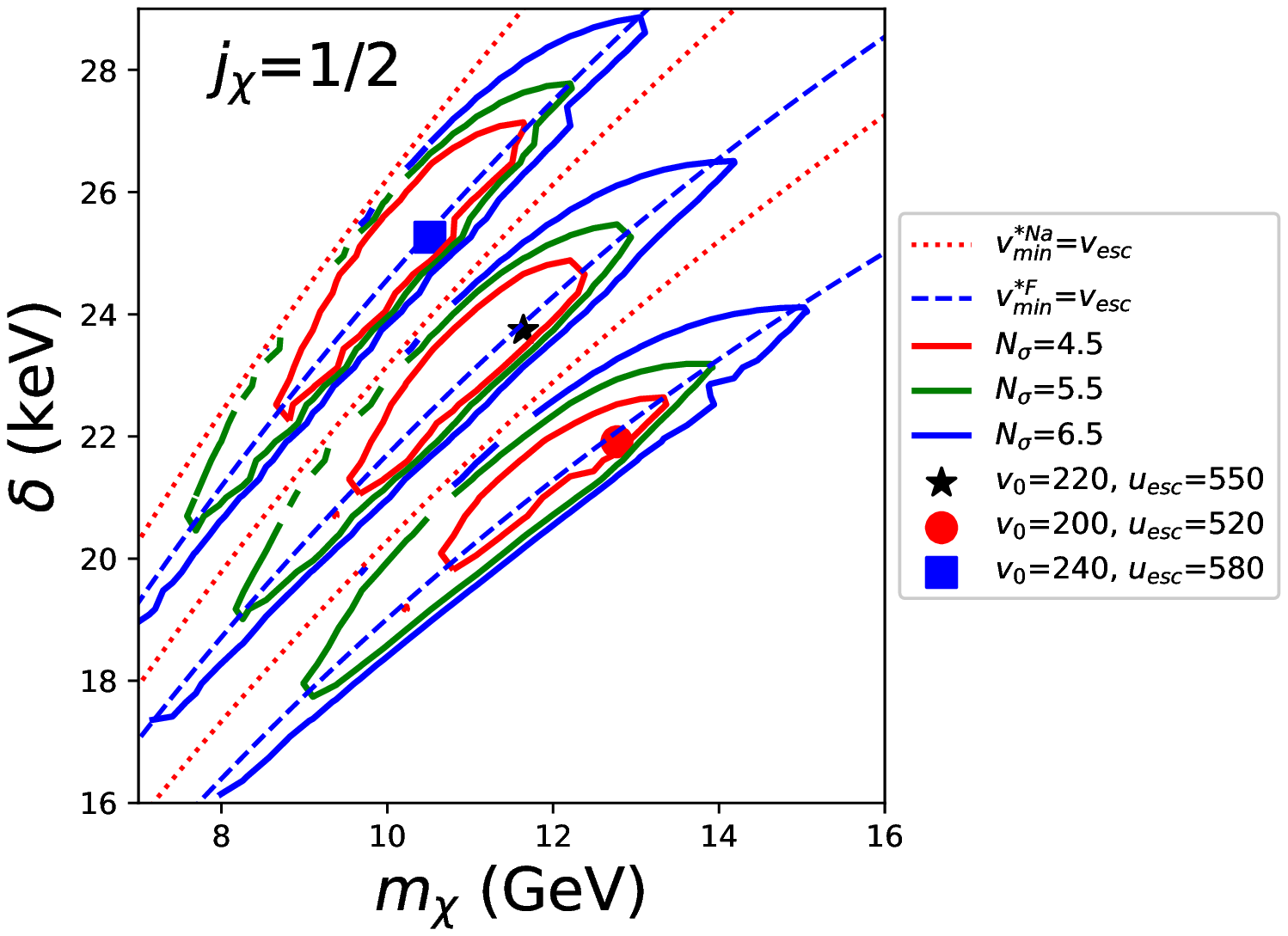}     
\end{center}
\caption{The same as Fig.~\ref{fig:mchi_delta_spin_0} for $j_{\chi}$=1/2.}
\label{fig:mchi_delta_spin_1_2}
\end{figure}

\begin{figure}
\begin{center}
\includegraphics[width=0.5\textwidth]{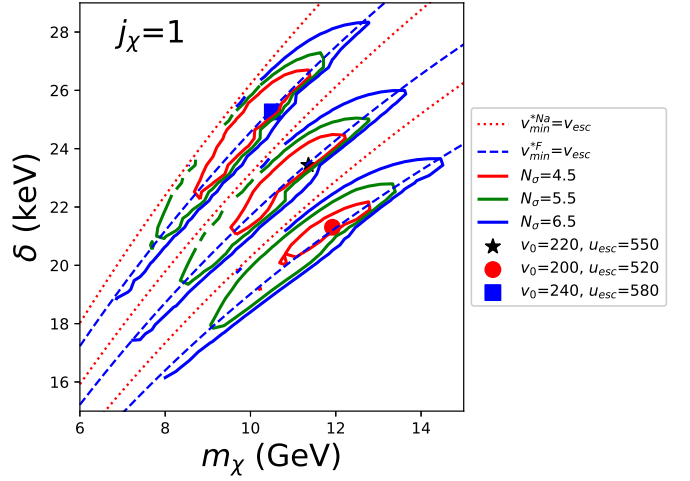}     
\end{center}
\caption{The same as Fig.~\ref{fig:mchi_delta_spin_0} for $j_{\chi}$=1.}
\label{fig:mchi_delta_spin_1}
\end{figure}
When the quantity $N_{\sigma}$ is plotted in the $m_{\chi}$--$\delta$
plane a general result common to all $j_{\chi}$ values is that the
region of parameter space where the tension is relieved is localized
in a narrow region with $\delta\gsim$ 20 keV. In
Figs.~\ref{fig:mchi_delta_spin_0},\ref{fig:mchi_delta_spin_1_2} and
\ref{fig:mchi_delta_spin_1} we provide contour plots of $N_{\sigma}$
centered on such localized regions of the $m_{\chi}$--$\delta$
parameter space for $j_{\chi}$=0, 1/2, 1, respectively, and for the
three combinations ($v_0$,$u_{esc}$)=(220,550),(200,520),(240,580)
km/s.  In all the plots the dotted (red) line represents the maximal
value of $\delta$ beyond which the minimal speed $v^{*Na}_{min}$
introduced in Eq.~(\ref{eq:vstar}) exceeds the escape velocity in the
Lab frame, $v_{esc}^{lab}$, while the (blue) short dashes show the
minimal value of $\delta$ beyond which
$v^{*F}_{min}<v_{esc}^{lab}$. This implies that between the two lines
the condition of Eq.~(\ref{eq:hierarchy}) is verified. The closed
contour where the tension $N_{\sigma}$ drops tracks for different
astrophysical parameters the region between the two lines, an
unequivocal indication that the same kinematic mechanism is at work as
in the pSIDM scenario~\cite{psidm_2015,psidm_2017,psidm_2018}
summarized in the Introduction. This also explains why, as observed in
Fig.~\ref{fig:lmi_mchi}, $N_{\sigma}$ is considerably more sensitive
on the astrophysical parameters $v_0$ and $u_{esc}$ compared to the
elastic case. So we conclude that the pSIDM scenario described
in~(\cite{psidm_2015,psidm_2017,psidm_2018}) emerges as the unique
mechanism to ease the tension between DAMA and other constraints from
a general scan of the inelastic DM parameter space.

In spite of the residual tension between DAMA and other constraints it
is interesting to discuss in detail the solutions corresponding to the
minimum values of $N_{\sigma}$.  In order to do so one needs to go
back to Eq.~(\ref{eq:lmi}).  At the boundary of its positivity the
smallest eigenvalue of the matrix in that equation is vanishing and the
corresponding eigenvector ${\bf \hat{c}}_{0}$ ($|{\bf \hat{c}}_{0}|$=1):

\begin{equation}
\left [\sum_{i\in\varmathbb{E}} \xi_i
  \varmathbb{A}_k-\frac{\varmathbb{S}_{m,n}^{DAMA}}{\hat{S}^{DAMA,min}_{m,n}}\right ]{\bf \hat{c}}_{0}\equiv \varmathbb{M}_{max}{\bf \hat{c}}_{0}=0,
\label{eq:eigen}
\end{equation}

\noindent individuates the line joining the origin to the points of
intersections between the extreme DAMA ellipsoid and those of the
constraints, as shown schematically by the solid line in
Fig.~\ref{fig:ellipses_example}. On the other hand, the vector ${\bf
  c}_{0,max}$, joining the origin to the intersection points and given
by:

\begin{equation}
{\bf c}_{0,max}=\sqrt{\frac{\hat{S}^{DAMA,max}_{m,n}}{{\bf \hat{c}}^T_0 \varmathbb{S}_{m,n}^{DAMA}\;{\bf \hat{c}}_0}} {\bf \hat{c}}_0
  \label{eq:c_max}
  \end{equation}

\noindent contains the set of couplings for the configuration of
maximal modulation amplitude in the bin $n$, and in
Fig.~\ref{fig:ellipses_example} is represented by the black arrow.
The properties of the specific point ${\bf c}_{0,max}$ in the space of
couplings can be further elucidated if one takes a closer look at the
matrices in Eqs.~(\ref{eq:matrices_sm},\ref{eq:matrices_s0}).
Depending on the spin of the WIMP particle they can have a different
dimensionality, as can be simply read--off from the WIMP response
functions in
Eqs.~(\ref{eq:wimp_response_functions},\ref{eq:wimp_response_functions_spin_1}).
Moreover, not all couplings interfere, so that the matrices can be
decomposed into block-diagonal form. The dimensionalities and
non--interfering subspaces are indicated in
Table~\ref{table:subspaces} for different values of the WIMP spin
$j_{\chi}$. This implies that also the matrix $\varmathbb{M}_{max}$ is
block--diagonal, so that ${\bf c}_{0,max}$ must belong to one of the
subspaces of Table~\ref{table:subspaces}.

\begin{table}[t]
\begin{center}
{\begin{tabular}{@{}|c|c|c|@{}}
\hline
spin &  couplings  &  dimensionality\\
\hline
0 &  $(c^{\tau}_1,c^{\tau}_3)$,$c^{\tau}_7$,$c^{\tau}_{10}$  &  2$\times (2+1+1)$=8\\
\hline
$\frac{1}{2}$ &  $(c^{\tau}_1,c^{\tau}_3)$,$(c^{\tau}_4,c^{\tau}_5,c^{\tau}_6)$, &2$\times(2+3+1+2+$  \\
& $c^{\tau}_7$, $(c^{\tau}_8,c^{\tau}_9)$, $c^{\tau}_{10}$, & $+1+3+1+1)$=28  \\
&    $(c^{\tau}_{11},c^{\tau}_{12},c^{\tau}_{15})$,$c^{\tau}_{13}$,$c^{\tau}_{14}$  &  \\
\hline
1  &   $c^{\tau}_1$,$(c^{\tau}_4,c^{\tau}_5)$, $(c^{\tau}_8,c^{\tau}_9)$,  &2$\times(1+2+2+1+$\\
     & $c^{\tau}_{10}$, $c^{\tau}_{11}$, $c^{\tau}_{14}$, $c^{\tau}_{17}$, $c^{\tau}_{18}$  & $1+1+1+1)$=20  \\
\hline
\end{tabular}}
\caption{Non--interfering subspaces and dimensionality of the coupling
  constants vector space of the NR WIMP effective theory of a WIMP
  with $j_{\chi}\le$1.
  \label{table:subspaces}}
\end{center}
\end{table}

\begin{table}[t]
\begin{center}
{\begin{tabular}{@{}|c|c|c|c|c|c|@{}}
\hline
spin &  $m_{\chi,0}$ (GeV)  & $\delta (keV)$ & $\sigma_{0,max}$ (cm$^2$)  & ${\bf \hat{c}}_{0}$ & $N_{\sigma}$\\
\hline
0 &  11.08  & 22.83 & 3.93$\times 10^{-27}$   & ($\hat{c}^0_{7}$=0.68, $\hat{c}^1_{7}$=0.73) & 4.0 \\
\hline
1/2 &  11.64 & 23.74 & 4.68$\times 10^{-28}$   & ($\hat{c}^0_{4}$=-0.0014, $\hat{c}^1_{4}$=-0.0015, & 2.9\\
&        &       &                &  $\hat{c}^0_{5}$=-0.032, $\hat{c}^1_{5}$=-0.0166, &\\
&        &        &               &  $\hat{c}^0_{6}$=0.692, $\hat{c}^1_{6}$=0.7217) &\\
\hline
1 &  11.36 & 23.43 & 5.71$\times 10^{-32}$   & ($\hat{c}^0_{4}$=0.0717, $\hat{c}^1_{4}$=0.0753,& 3.2 \\
 &        &      &                 & $\hat{c}^0_{5}$=0.1892, $\hat{c}^1_{5}$=0.9764) & \\
\hline
\end{tabular}}
\caption{Properties of the extreme configurations in the NR effective
  theory parameter space that minimize the tension $N_{\sigma}$ for
  different values of the WIMP spin $j_{\chi}$ and fixing the
  astrophysocal parameters to $v_0$=220 km/s and $u_{esc}$=550 km/s.
  \label{table:extremes}}
\end{center}
\end{table}

The properties of the extreme configurations found in this way are
given in Table~\ref{table:extremes}. Interestingly, the configuration
with the smallest tensions corresponds to a $c_{7}$ coupling for
$j_{\chi}$=0 (a spin--dependent interaction with explicit velocity
dependence and momentum suppression $q^2$)) and to approximately a
$c_6$ coupling for $j_{\chi}$=1/2 (a spin--dependent interaction with
momentum suppression $q^4$). These two couplings combinations
correspond to two of the possible generalizations of the pSIDM
scenario already discussed in~\cite{psidm_2017}. On the other hand,
for $j_{\chi}$=1 the extreme configuration corresponds to a dominant
$c_5$ coupling (associated to the WIMP coupling to the orbital angular
momentum operator) also with momentum suppression $q^4$, and a
non--negligible $c_4$ contribution.  The role of momentum suppression
in relieving the tension between the DAMA result and other constraints
has already been pointed out in~\cite{eft_spin}.

Our procedure does not correspond to the minimization of a
$\chi$--square, since we minimize the tension in one target bin at a
time. However, once a minimal tension configuration is obtained, the
quantity $\chi^2$=$\sum
[S^{DAMA}_{m,n}-S^{DAMA}_{m,n,exp}]^2/\sigma_{exp}^2$ (with
$S^{DAMA}_{m,n,exp}$ and $\sigma_{exp}$ the measured modulation
amplitudes and standard deviations) can be calculated. In this way,
for the configurations of Table~\ref{table:extremes} we get
$\chi^2$=60.6, 25.7 and 33.5 for $j_{\chi}$=0,1/2 and 1, respectively.
\begin{figure}
\begin{center}
\includegraphics[width=0.5\textwidth]{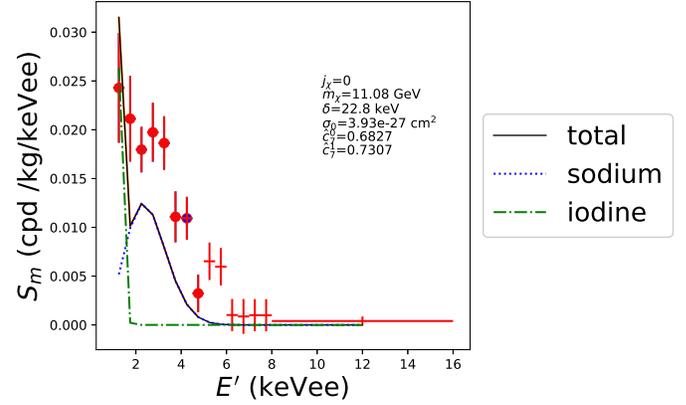}     
\end{center}
\caption{Predictions of the DAMA modulation amplitudes for the
  configuration of minimal $N_{\sigma}$ and $j_{\chi}$=0 shown in
  Table~\ref{table:extremes} vs. the DAMA experimental
  measurements. Experimental intervals represent the combination of
  DAMA/LIBRA--phase1 and DAMA/LIBRA--phase2 from~\cite{dama_2018}. The
  (black) solid line represents the predicted modulation amplitudes for $NaI$,
  while the (green) dot--dashed line (vanishing above 2 keVee) and (blue)
  dotted line show the separate contributions from WIMP scattering
  events off Iodine and Sodium, respectively. The experimental points
  marked with a (red) circle correspond to the energy bins included in
  the solution of Eq.~(\ref{eq:lmi}), while the point marked with an
  additional (blue) inner circle corresponds to the DAMA energy bin
  where the maximal tension with the bounds arises and that drives the
  determination of $N_{\sigma}$. in Figs.~\ref{fig:lmi_mchi} and
  \ref{fig:mchi_delta_spin_0}.}
\label{fig:e_sm_spin_0}
\end{figure}

\begin{figure}
\begin{center}
\includegraphics[width=0.5\textwidth]{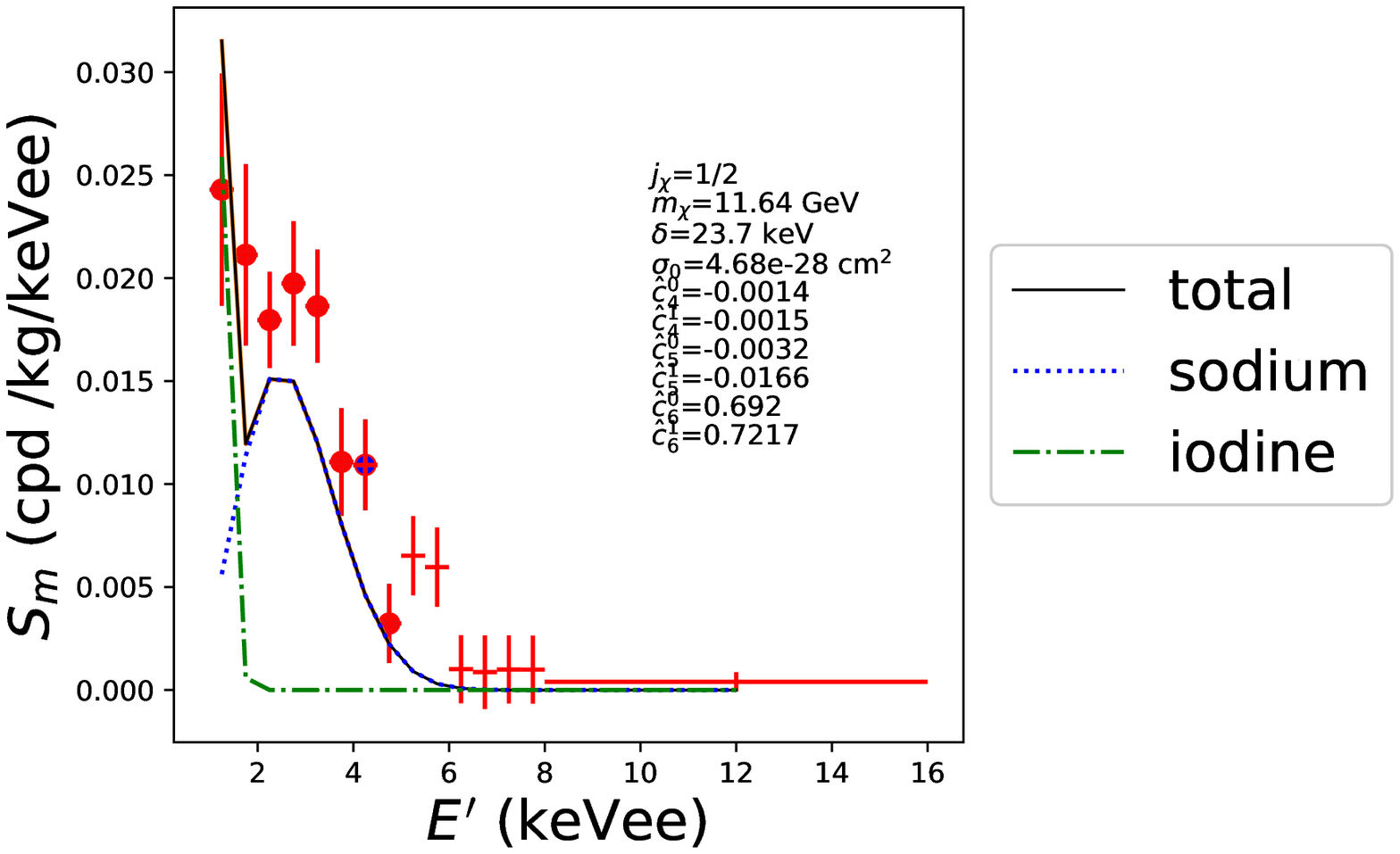}     
\end{center}
\caption{The same as in Fig.~\ref{fig:e_sm_spin_0} for $j_{\chi}$=1/2.}
\label{fig:e_sm_spin_1_2}
\end{figure}

\begin{figure}
\begin{center}
\includegraphics[width=0.5\textwidth]{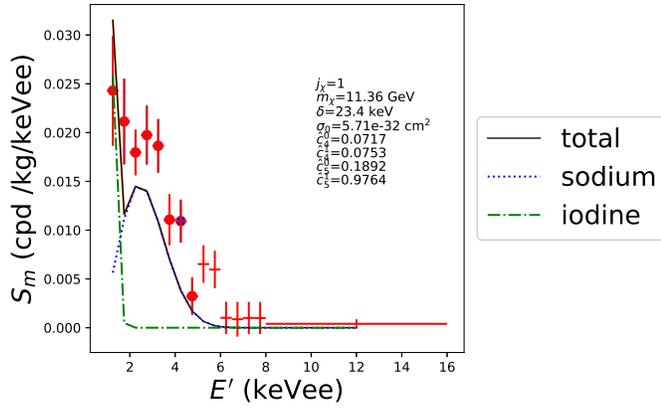}     
\end{center}
\caption{The same as in Fig.~\ref{fig:e_sm_spin_0} for $j_{\chi}$=1.}
\label{fig:e_sm_spin_1}
\end{figure}

The corresponding predictions for the DAMA modulation amplitudes are
shown in Figs.~\ref{fig:e_sm_spin_0},\ref{fig:e_sm_spin_1_2} and
\ref{fig:e_sm_spin_1} for the different values of $j_{\chi}$ and
compared to the measured ones~\cite{dama_2018}. In such figures the
experimental points marked with a (red) circle correspond to the
energy bins included in the solution of Eq.~(\ref{eq:lmi}), while the
point marked with an additional (blue) inner circle corresponds to the
DAMA energy bin where the maximal tension with the bounds arises and
that drives the determination of $N_{\sigma}$.  As one can see in the
lower energy bins the allowed modulation signal is large enough to
explain the DAMA signal, although the amplitudes spectrum decays
faster with energy.

\begin{figure}
\begin{center}
\includegraphics[width=0.5\textwidth]{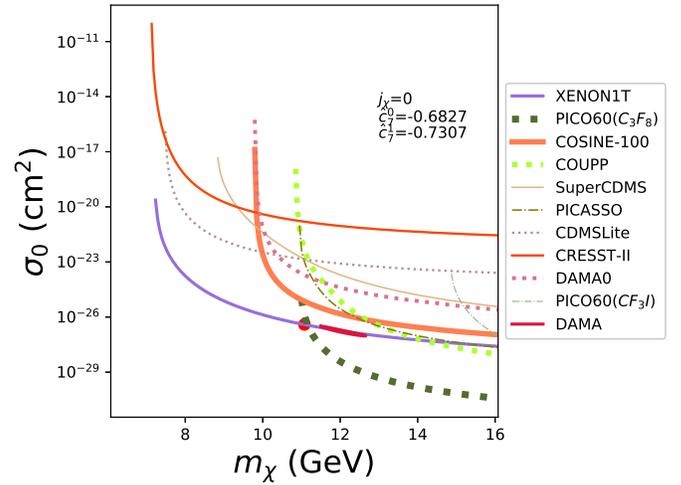}     
\end{center}
\caption{Experimental upper bounds (open lines) and 2$\sigma$ DAMA
  region (closed contour) in the $m_{\chi}$--$\sigma_0$ parameter
  space for $j_{\chi}$=0. In this figure the mass splitting $\delta$
  and the direction in coupling ${\bf \hat{c}}_0$ are fixed to the
  values of Table~\ref{table:extremes}, while the effective cross
  section $\sigma_0$ is defined in Eq.~(\ref{eq:sigma_0}).The (red)
  circle represents the point in parameter space with minimal
  $N_{\sigma}$.\label{fig:mchi_sigma_spin_0}}
\end{figure}

\begin{figure}
\begin{center}
\includegraphics[width=0.5\textwidth]{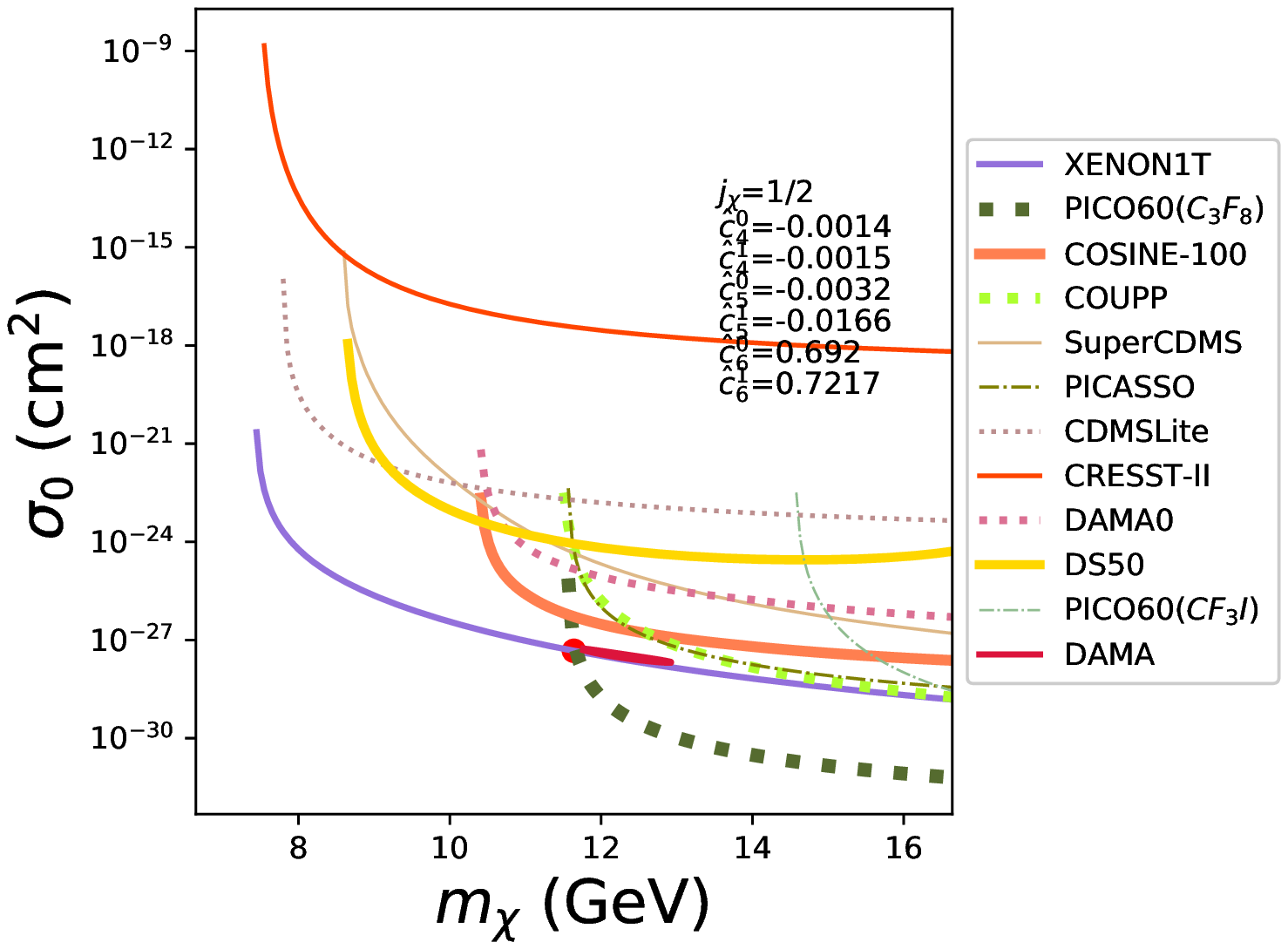}     
\end{center}
\caption{The same as in Fig.~\ref{fig:mchi_sigma_spin_0} for $j_{\chi}$=1/2.}
\label{fig:mchi_sigma_spin_1_2}
\end{figure}

\begin{figure}
\begin{center}
\includegraphics[width=0.5\textwidth]{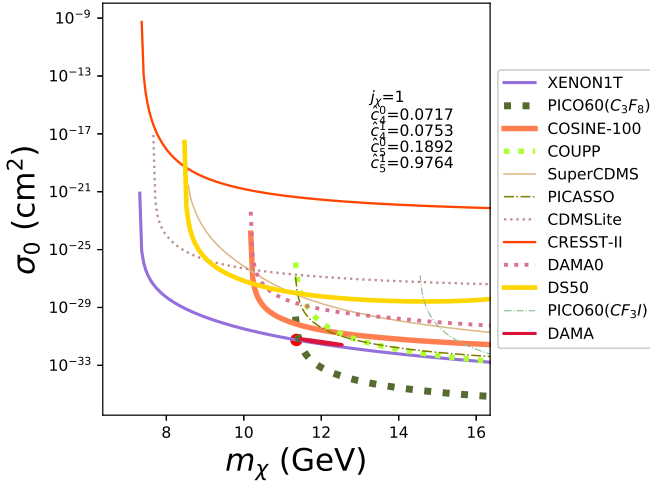}     
\end{center}
\caption{The same as in Fig.~\ref{fig:mchi_sigma_spin_0} for $j_{\chi}$=1.}
\label{fig:mchi_sigma_spin_1}
\end{figure}

Introducing the vector ${\bf c}_0\equiv c_0 {\bf \hat{c}}_0$ with
$c_0$ a free normalization, $c_0$ is common to all signals inside the
eigenspace of ${\bf \hat{c}}_0$.  A convenient parameterization is
through the introduction of the reference cross section:

\begin{equation}
\sigma_{0}\equiv c_0^2\frac{\mu_{\chi{\cal N}}^2}{\pi},
  \label{eq:sigma_0}
  \end{equation}
\noindent with $\mu_{\chi{\cal N}}$ the WIMP--nucleon reduced mass.
The direction in coupling space singled out by the unit vector ${\bf
  \hat{c}}_0$, and that individuates a specific set of coupling ratios
that eases the tension between DAMA and the constraints, can be seen
as the generalization in an arbitrary number of dimensions of the
concept of isospin--violating DM~\cite{isospin_violation}. Once ${\bf
  \hat{c}}_0$ and $\delta$ are fixed the DAMA signal and the bounds
can be discussed in a familiar $m_{\chi}$--$\sigma_{0}$ plane. This is
done in Figs.~\ref{fig:mchi_sigma_spin_0},
\ref{fig:mchi_sigma_spin_1_2} and \ref{fig:mchi_sigma_spin_1} for the
extreme configurations summarized in Table~\ref{table:extremes} and
for $j_{\chi}$=1, 1/2 and 1, respectively. In the same figure the
(red) circle represents the point $(m_{\chi,0},\sigma_{0,max})$ with
$m_{\chi,0}$ the value of the WIMP mass for the extreme configuration
and $\sigma_{0,max}=c^2_{0,max} \frac{\mu_{\chi{\cal N}}^2}{\pi}$,
$c_{0,max}\equiv |{\bf c}_{0,max}|$. The point
$(m_{\chi,0},\sigma_{0,max})$ intersects one or more of the most
constraining boundaries on $\sigma_0$ at $m_{\chi}$=$m_{\chi,0}$,
providing a nice confirmation of the numerical solution of
Eq.~(\ref{eq:lmi}). Actually, this can be directly observed in
Figs.~\ref{fig:mchi_sigma_spin_0}, \ref{fig:mchi_sigma_spin_1_2} and
\ref{fig:mchi_sigma_spin_1} for for $j_{\chi}$=0, 1/2 and 1,
respectively, where the point $(m_{\chi,0},\sigma_{0,max})$ lies on
the intersection between the bounds from XENON1T and PICO--60, in a
realization of the mechanism shown schematically in
Fig.~\ref{fig:ellipses_example}. In all three cases, the extreme
configuration hits also the 90\% C.L. upper bound on the modulation
fraction in the first bin, as shown in Figs.~\ref{fig:e_sm_spin_0},
\ref{fig:e_sm_spin_1_2} and \ref{fig:e_sm_spin_1}. In
Figs.~\ref{fig:mchi_sigma_spin_0}, ~\ref{fig:mchi_sigma_spin_1_2} and
~\ref{fig:mchi_sigma_spin_1} the distance from the DAMA region and the
extreme point provides an additional visual indicator besides
$N_{\sigma}$ of the tension between DAMA and the constraints from null
results. Indeed, as already observed in \cite{psidm_2018}, the
condition~(\ref{eq:hierarchy}) implies that inside the energy range of
the DAMA effect the spectrum of the predicted modulation amplitudes
has a maximum corresponding to the recoil energy $E_R^*\equiv
E_R(v^{*Na}_{min})$=$|\delta|\mu_{\chi N}/m_N$ for scattering events
off sodium. On the other hand, the data from DAMA/LIBRA-phase2 are
more compatible to a monotonically decreasing shape\footnote{The
  DAMA/LIBRA-phase1 data showed instead a maximum, and for this reason
  inelastic scattering could provide a good fit to the
  data~\cite{psidm_2015}.} closer to elastic scattering. As a
consequence, the DAMA data pull to low values of $\delta$. However,
the solutions of Eq.~(\ref{eq:lmi}) with smallest tension with the
constraints require sizeable values of $\delta$ ($\delta\gsim$ 20keV)
in order to verify Eq.~(\ref{eq:hierarchy}). As a consequence, when
$\delta$ is fixed to such values the DAMA data pull to higher values
of the WIMP mass $m_{\chi}$ to dilute the effect of $\delta$. This
explains why, systematically, the DAMA regions in
Figs.~\ref{fig:mchi_sigma_spin_0}, ~\ref{fig:mchi_sigma_spin_1_2} and
~\ref{fig:mchi_sigma_spin_1} are at higher WIMP masses compared to the
values of $m_{\chi,0}$ in Table~\ref{table:extremes}.  Moreover,
Figs.~\ref{fig:e_sm_spin_0}, \ref{fig:e_sm_spin_1_2} and
\ref{fig:e_sm_spin_1} show that configurations allowed by constraints
from null results can produce enough yearly modulation in some of the
DAMA bins, but the ensuing shape of the modulation spectrum is too
steep, so that the maximal modulation at high energy is constrained by
the bins at low energy. In light of this observation, we interpret the
fact that all the smallest tension configurations of
Table~\ref{table:extremes} have an interaction with explicit momentum
suppression as a way to alleviate this problem by suppressing the DAMA
modulation amplitudes in the lowest--energy bins.

These findings
are in agreements to those of Ref.~\cite{psidm_2018}, obtained for the
specific case of a standard spin--dependent interaction.

Equation (\ref{eq:lmi}) can only be solved for a limited selection of
experimental bounds both because of computing time limits, and because
some of the constraints require more refined treatments beside a
simple comparison between theoretical predictions and upper bounds as
in Eq.~(\ref{eq:upper_bounds}) and Table~\ref{table:exp_lmi}, such as
background subtraction or the optimal-interval
method~\cite{yellin}. In Figs.~\ref{fig:mchi_sigma_spin_0},
\ref{fig:mchi_sigma_spin_1_2} and \ref{fig:mchi_sigma_spin_1} all this
standard machinery~\cite{sensitivities_2018} can instead be applied to
the full set of existing experiments, providing an {\it a posteriori}
confirmation that the set of bounds $\varmathbb{E}$ used to solve
Eq.~(\ref{eq:lmi}) did not miss any relevant constraint.  In
particular, besides the experiments included in the solution of
Eq.~(\ref{eq:lmi}), in such figures we have added
CDMSlite~\cite{cdmslite_2017},
CRESST-II~\cite{cresst_II,cresst_II_ancillary}, the upper bound from
the average count rate of DAMA~\cite{damaz}), DarkSide--50~\cite{ds50}
and the $CF_3I$ target run of PICO--60~\cite{pico60_2015} (the details
of such bounds implementation are provided in Appendix \ref{app:exp}).
None of these additional null results further constrains the extreme
configurations ($m_{\chi,0},\sigma_{0,max}$).

We conclude by noting that the recent bound from
COSINE--100~\cite{cosine_nature}, obtained with the same $NaI$ target
material as DAMA, is not particularly binding in our analysis, as can
be seen again in Figs.~\ref{fig:mchi_sigma_spin_0},
\ref{fig:mchi_sigma_spin_1_2} and \ref{fig:mchi_sigma_spin_1}.  The
reason of this is that the bound of Ref.~\cite{cosine_nature} is on
the time--averaged signal $S_0^{COSINE}$, while DAMA measures the
yearly modulation amplitudes expected from the rotation of the Earth
around the Sun $S_m^{DAMA}$. In appendix~\ref{app:exp} we estimate in
COSINE--100 a residual count rate $b\simeq$0.13 events/kg/day/keV
after background subtraction while modulation fractions in DAMA are of
the order of 0.02 events/kg/day/keV. The bound $S_0^{COSINE}<b$
implies $S_m^{DAMA}/S_0^{DAMA}$ = $S_m^{DAMA}/S_0^{COSINE}\times
S_0^{COSINE}/S_0^{DAMA}$ $\gsim$ 0.12, including a factor
$S_0^{COSINE}/S_0^{DAMA}\simeq$ 0.8 due to a difference between the
energy resolutions and efficiencies in the two experiments. For a
standard Maxwellian WIMP velocity distribution in the SI elastic case
the predicted modulation fractions $S_m^{DAMA}/S_0^{DAMA}$ are below
such bound (for instance, for $m_{\chi}$=10 GeV
$S_m^{DAMA}/S_0^{DAMA}$ is between $\simeq$0.05 and $\simeq$0.12 for
$E_{ee}<$ 3 keVee) explaining why in Ref.~\cite{cosine_nature} the
DAMA effect is ruled out. However expected rates for inelastic
scattering are sensitive to the high--speed tail of the WIMP velocity
distribution for which the modulation fractions are sizeably
higher~\cite{inelastic, psidm_2015}, and this is particularly true when the
condition~(\ref{eq:hierarchy}) is verified (for instance, in the
extreme configurations of Table~\ref{table:extremes} we find
$S_m^{DAMA}/S_0^{DAMA}\gsim$0.8 for $E^{\prime}\le$3.5 keVee).

\section{Conclusions}
\label{sec:conclusions}

In the present paper we have discussed the compatibility of the
combined annual modulation effect measured by DAMA/LIBRA--phase1 and
DAMA/LIBRA--phase2~\cite{dama_2008,dama_2010,dama_2013,dama_2018} with
an explanation in terms of inelastic scattering events induced by the
most general Galilean-invariant effective contact interaction of a
spin 0, 1/2 or 1 WIMP dark matter particle taking into account all the
possible interferences among operators by studying the intersections
among the ellipsoidal surfaces of constant signal of DAMA and other
experiments in the space of the coupling constants of the effective
theory, following the approach introduced in Ref.~\cite{Catena_dama}.
In our analysis we have assumed a standard Maxwellian velocity
distribution in the Galaxy.  Compared to the elastic case, inelastic
scattering partially relieves but does not eliminate the existing
tension between the DAMA effect and the constraints from the null
results of other experiments. We have determined the ellipsoids using
90\% C.L. upper bounds from selected energy bins from
XENON1T~\cite{xenon_2018}, PICO--60 ($C_3F_8$ target) \cite{pico60,pico60_2019},
COSINE--100\cite{cosine_nature}, COUPP~\cite{coupp},
SuperCDMS~\cite{super_cdms_2017} and PICASSO~\cite{picasso}.  The
tension, quantified as the maximum of
$N_{\sigma}\equiv\max(n_{\sigma})$ among the DAMA energy bins below 5
keVee, exceeds 7$\sigma$ in all the parameter space $m_{\chi}<$ 200
GeV with the exception of a small region of parameter space for
$m_{\chi}\simeq$ 10 GeV and $\delta\gsim$20, where it drops to values
as low as $\simeq$ 4$\sigma$ for $j_{\chi}$=0, $\simeq$ 2.9$\sigma$
for $j_{\chi}$=1/2 and $\simeq$ 3.2$\sigma$ for $j_{\chi}$=1, and
that overlaps to the proton--philic spin--dependent inelastic Dark
Matter (pSIDM) scenario~\cite{psidm_2015,psidm_2017,psidm_2018}
already discussed in the literature for the specific case of a
standard spin--dependent interaction, where the bounds from fluorine
targets are evaded in a kinematic way because the minimal WIMP
incoming speed required to trigger upscatters off fluorine exceeds the
maximal WIMP velocity in the Galaxy, or is very close to it.  In
particular, from a general scan of the inelastic DM parameter space
such kinematic feature, together with momentum suppression in the
effective operator, emerge as instrumental in easing the tension
between DAMA and other constraints. As a consequence, the latter is
considerably more sensitive on the astrophysical parameters compared
to the elastic case.  The configurations for which the tension
$N_{\sigma}$ is partially relieved can easily produce enough yearly
modulation in the lowest--energy bins of the modulation spectrum
measured by DAMA in compliance with the constraints from other
experiments. However, the ensuing shape of the modulation spectrum is
too steep, so that, when not excluded by other constraints, the
maximal allowed modulation at higher energies is constrained by the
modulation measured in the lowest energy bins.

The present analysis extends the scope of previous ones in the task to
explore a DAMA explanation in the full WIMP direct detection parameter
space, but is still not the most general one. Possible extensions
include: i) long--range interactions; ii) allowing for complex
couplings; iii) assuming a WIMP velocity distribution that departs
from a standard Maxwellian. In particular, given the large dependence
on the astrophysical parameters that we observed in our results we
expect the latter generalization as very promising in order to find
effective models that reconcile the DAMA result with the null
observations of other experiments.

\acknowledgments This research was supported by the Basic Science
Research Program through the National Research Foundation of Korea
(NRF) funded by the Ministry of Education, grant number
2016R1D1A1A09917964. 

\section*{Note added}
After the submission of the present paper COSINE--100 has released its
first annual modulation analysis, consistent at 68.3\% C.L.  with both
a null hypothesis and DAMA/LIBRA's 2--6 keVee best--fit
value~\cite{cosine_2019_modulation}.

\appendix

\section{WIMP response functions}
\label{app:wimp_eft}

We collect here the WIMP particle--physics response functions
introduced in Eq.(\ref{eq:squared_amplitude}) and for the general case
of complex couplings~\cite{eft_inelastic} (although in the present
analysis real couplings have been assumed). For a WIMP particle of
spin $J_{\chi}\le\frac{1}{2}$ they are given
by~\cite{haxton1,haxton2}:

\begin{eqnarray}
&& R_{M}^{\tau \tau^\prime}\left(v_T^{\perp 2}, {q^2 \over m_N^2}\right) = c_1^\tau c_1^{\tau^\prime *} + {j_\chi (j_\chi+1) \over 3} \left[ {q^2 \over m_N^2} v_T^{\perp 2} c_5^\tau c_5^{\tau^\prime * }+\nonumber\right .\\
&&\left. +v_T^{\perp 2}c_8^\tau c_8^{\tau^\prime * } + {q^2 \over m_N^2} c_{11}^\tau c_{11}^{\tau^\prime * } \right], \nonumber \\
  && R_{\Phi^{\prime \prime}}^{\tau \tau^\prime}\left(v_T^{\perp 2}, {q^2 \over m_N^2}\right) = \left [{q^2 \over 4 m_N^2} c_3^\tau c_3^{\tau^\prime * } + {j_\chi (j_\chi+1) \over 12} \left( c_{12}^\tau-{q^2 \over m_N^2} c_{15}^\tau\right)\right .\nonumber\\
&&\left .\times\left( c_{12}^{\tau^\prime * }-{q^2 \over m_N^2}c_{15}^{\tau^\prime *} \right)\right ]\frac{q^2}{m_N^2},  \nonumber \\
  && R_{\Phi^{\prime \prime} M}^{\tau \tau^\prime}\left(v_T^{\perp 2}, {q^2 \over m_N^2}\right) =\mbox{Re} \left [ c_3^\tau c_1^{\tau^\prime * }+\right.\nonumber\\
  &&  \left .+ {j_\chi (j_\chi+1) \over 3} \left( c_{12}^\tau -{q^2 \over m_N^2} c_{15}^\tau \right) c_{11}^{\tau^\prime * }\right ] \frac{q^2}{m_N^2}, \nonumber \\
&&  R_{\tilde{\Phi}^\prime}^{\tau \tau^\prime}\left(v_T^{\perp 2}, {q^2 \over m_N^2}\right) =\left [{j_\chi (j_\chi+1) \over 12} \left ( c_{12}^\tau c_{12}^{\tau^\prime * }+{q^2 \over m_N^2}  c_{13}^\tau c_{13}^{\tau^\prime *}  \right )\right ]\frac{q^2}{m_N^2}, \nonumber \\
&&   R_{\Sigma^{\prime \prime}}^{\tau \tau^\prime *}\left(v_T^{\perp 2}, {q^2 \over m_N^2}\right) ={q^2 \over 4 m_N^2} c_{10}^\tau  c_{10}^{\tau^\prime * } +
  {j_\chi (j_\chi+1) \over 12} \left[ c_4^\tau c_4^{\tau^\prime *} + \right.  \nonumber \\
 && \left. +{q^2 \over m_N^2} ( c_4^\tau c_6^{\tau^\prime * }+c_6^\tau c_4^{\tau^\prime * })+
    {q^4 \over m_N^4} c_{6}^\tau c_{6}^{\tau^\prime * } +v_T^{\perp 2} c_{12}^\tau c_{12}^{\tau^\prime * }+\right .\nonumber\\
    &&\left.+{q^2 \over m_N^2} v_T^{\perp 2} c_{13}^\tau c_{13}^{\tau^\prime * } \right], \nonumber \\
   && R_{\Sigma^\prime}^{\tau \tau^\prime}\left(v_T^{\perp 2}, {q^2 \over m_N^2}\right)  ={1 \over 8} \left[ {q^2 \over  m_N^2}  v_T^{\perp 2} c_{3}^\tau  c_{3}^{\tau^\prime * } + v_T^{\perp 2}  c_{7}^\tau  c_{7}^{\tau^\prime * }  \right]+\nonumber\\
&&+ {j_\chi (j_\chi+1) \over 12} \left[ c_4^\tau c_4^{\tau^\prime *} +{q^2 \over m_N^2} c_9^\tau c_9^{\tau^\prime * }+  \right.\nonumber \\
    &&\left. +{v_T^{\perp 2} \over 2} \left(c_{12}^\tau-{q^2 \over m_N^2}c_{15}^\tau \right) \left( c_{12}^{\tau^\prime * }-{q^2 \over m_N^2}c_{15}^{\tau \prime *} \right)+{q^2 \over 2 m_N^2} v_T^{\perp 2}  c_{14}^\tau c_{14}^{\tau^\prime * } \right], \nonumber \\
    && R_{\Delta}^{\tau \tau^\prime}\left(v_T^{\perp 2}, {q^2 \over m_N^2}\right)= {j_\chi (j_\chi+1) \over 3} \left( {q^2 \over m_N^2} c_{5}^\tau c_{5}^{\tau^\prime * }+ c_{8}^\tau c_{8}^{\tau^\prime * } \right)\frac{q^2}{m_N^2}, \nonumber \\
  && R_{\Delta \Sigma^\prime}^{\tau \tau^\prime}\left(v_T^{\perp 2}, {q^2 \over m_N^2}\right)= {j_\chi (j_\chi+1) \over 3} \mbox{Re}\left (c_{5}^\tau c_{4}^{\tau^\prime * }-c_8^\tau c_9^{\tau^\prime *} \right) \frac{q^2}{m_N^2}.\label{eq:wimp_response_functions}
\end{eqnarray}
\noindent  On the other hand, for a WIMP particle with spin $J_{\chi}$=1~\cite{krauss_spin_1}:
\begin{eqnarray}
&& R_{M}^{\tau \tau^\prime}\left(v_T^{\perp 2}, {q^2 \over m_N^2}\right) = c_1^\tau c_1^{\tau^\prime * } + {2 \over 3} \left[ {q^2 \over m_N^2} v_T^{\perp 2} c_5^\tau c_5^{\tau^\prime * }+v_T^{\perp 2}c_8^\tau c_8^{\tau^\prime * }+\nonumber\right .\\
&&\left. + {q^2 \over m_N^2} c_{11}^\tau c_{11}^{\tau^\prime * }+{q^2\over {4 m_N^2}}v_T^{\perp 2}c_{17}^\tau c_{17}^{\tau^\prime * } \right], \nonumber \\
  && R_{\Phi^{\prime \prime}}^{\tau \tau^\prime}\left(v_T^{\perp 2}, {q^2 \over m_N^2}\right) =0,\nonumber\\
&& R_{\Phi^{\prime \prime} M}^{\tau \tau^\prime}\left(v_T^{\perp 2}, {q^2 \over m_N^2}\right) =0, \nonumber \\
&&  R_{\tilde{\Phi}^\prime}^{\tau \tau^\prime}\left(v_T^{\perp 2}, {q^2 \over m_N^2}\right) =0, \nonumber \\
&&   R_{\Sigma^{\prime \prime}}^{\tau \tau^\prime}\left(v_T^{\perp 2}, {q^2 \over m_N^2}\right) ={q^2 \over 4 m_N^2} c_{10}^\tau  c_{10}^{\tau^\prime * } +
  {1 \over 6}  c_4^\tau c_4^{\tau^\prime *} + {q^2\over {12 m_N^2}}v_T^{\perp 2}c_{18}^\tau c_{18}^{\tau^\prime * },\nonumber\\
  && R_{\Sigma^\prime}^{\tau \tau^\prime}\left(v_T^{\perp 2}, {q^2 \over m_N^2}\right)  = {1 \over 6} c_4^\tau c_4^{\tau^\prime *} +{q^2 \over m_N^2} c_9^\tau c_9^{\tau^\prime * }+{q^2 \over 2 m_N^2} v_T^{\perp 2}  c_{14}^\tau c_{14}^{\tau^\prime * }+\nonumber\\
  &&+{q^2\over {24 m_N^2}}c_{18}^\tau c_{18}^{\tau^\prime * } , \nonumber \\
    && R_{\Delta}^{\tau \tau^\prime}\left(v_T^{\perp 2}, {q^2 \over m_N^2}\right)= {2 \over 3} \left( {q^2 \over m_N^2} c_{5}^\tau c_{5}^{\tau^\prime * }+ c_{8}^\tau c_{8}^{\tau^\prime * } \right)\frac{q^2}{m_N^2}, \nonumber \\
&& R_{\Delta \Sigma^\prime}^{\tau \tau^\prime}\left(v_T^{\perp 2}, {q^2 \over m_N^2}\right)= {2 \over 3}\mbox{Re} \left (c_{5}^\tau c_{4}^{\tau^\prime * }-c_8^\tau c_9^{\tau^\prime *} \right) \frac{q^2}{m_N^2}.
  \label{eq:wimp_response_functions_spin_1}
\end{eqnarray}

\section{Experiments}
\label{app:exp}

\begin{table}[t]
\begin{center}
{\begin{tabular}{@{}|c|c|c|@{}}
\hline
Experiment &  visible energy range  &  90\% C.L. upper bound\\
\hline
DAMA & 1 keVee$<E^{\prime}<$1.5 keVee  &  0.0315\,\mbox{kg$^{-1}$day$^{-1}$keVee$^{-1}$}  \\
& 1.5 keVee$<E^{\prime}<$2 keVee  &  0.0268\,\mbox{kg$^{-1}$day$^{-1}$keVee$^{-1}$}  \\
& 2 keVee$<E^{\prime}<$2.5 keVee  &  0.0210\,\mbox{kg$^{-1}$day$^{-1}$keVee$^{-1}$}  \\
& 2.5 keVee$<E^{\prime}<$3 keVee  &  0.0236\,\mbox{kg$^{-1}$day$^{-1}$keVee$^{-1}$}  \\
& 3 keVee$<E^{\prime}<$3.5 keVee  &  0.0222\,\mbox{kg$^{-1}$day$^{-1}$keVee$^{-1}$}  \\
& 3.5 keVee$<E^{\prime}<$4 keVee  &  0.0144\,\mbox{kg$^{-1}$day$^{-1}$keVee$^{-1}$} \\
& 4 keVee$<E^{\prime}<$4.5 keVee  &  0.0137\,\mbox{kg$^{-1}$day$^{-1}$keVee$^{-1}$} \\
& 4.5 keVee$<E^{\prime}<$5 keVee  &  0.00569\,\mbox{kg$^{-1}$day$^{-1}$keVee$^{-1}$} \\
\hline
XENON1T & 3PE $<S_1<$70PE  &  11.77\,\mbox{events} \\
\hline
PICO--60 & $E_R>$2.45 keVnr  &  6.42\,\mbox{events} \\
         & $E_R>$3.3 keVnr  &  2.3\,\mbox{events} \\
\hline
COSINE--100& 2 keVee$<E^{\prime}<$2.5 keVee & 0.13\,\mbox{kg$^{-1}$day$^{-1}$keVee$^{-1}$} \\
           &  4.5 keVee$<E^{\prime}<$5 keVee & 0.13\,\mbox{kg$^{-1}$day$^{-1}$keVee$^{-1}$} \\
&  7.5 keVee$<E^{\prime}<$8 keVee & 0.13\,\mbox{kg$^{-1}$day$^{-1}$keVee$^{-1}$} \\
\hline
COUPP& $E_R>$7.8 keVnr & 6.68 \,\mbox{events} \\
     & $E_R>$11 keVnr & 5.32 \,\mbox{events} \\
& $E_R>$15.5 keVnr & 11.6 \,\mbox{events} \\
\hline
SuperCDMS&  4 keVnr$<E_R<$100 keVnr  &  3.89\,\mbox{events} \\
\hline
PICASSO& $E_R>$1.0 keVnr & 3.45\,\mbox{kg$^{-1}$day$^{-1}$keVee$^{-1}$} \\
\hline
\hline
Experiment &  visible energy range  &  90\% C.L. lower bound\\
\hline
DAMA & 1 keVee$<E^{\prime}<$1.5 keVee  & 0.0171\,\mbox{kg$^{-1}$day$^{-1}$keVee$^{-1}$}        \\
& 1.5 keVee$<E^{\prime}<$2 keVee  & 0.0155\,\mbox{kg$^{-1}$day$^{-1}$keVee$^{-1}$}            \\
& 2 keVee$<E^{\prime}<$2.5 keVee  & 0.0150\,\mbox{kg$^{-1}$day$^{-1}$keVee$^{-1}$}     \\
& 2.5 keVee$<E^{\prime}<$3 keVee  & 0.0159\,\mbox{kg$^{-1}$day$^{-1}$keVee$^{-1}$}     \\
& 3 keVee$<E^{\prime}<$3.5 keVee  & 0.0151\,\mbox{kg$^{-1}$day$^{-1}$keVee$^{-1}$}       \\
& 3.5 keVee$<E^{\prime}<$4 keVee  & 0.00773\,\mbox{kg$^{-1}$day$^{-1}$keVee$^{-1}$}       \\
& 4 keVee$<E^{\prime}<$4.5 keVee  & 0.00812\,\mbox{kg$^{-1}$day$^{-1}$keVee$^{-1}$}       \\
& 4.5 keVee$<E^{\prime}<$5 keVee  & 0.000770\,\mbox{kg$^{-1}$day$^{-1}$keVee$^{-1}$}       \\
\hline
\end{tabular}}
\caption{Visible energy intervals and 90\% C.L. upper or lower bounds
 used to calculate the matrices $\varmathbb{A}_{j}$ and $\varmathbb{B}_{j}$ in Eq.~(\ref{eq:lmi}).
  \label{table:exp_lmi}}
\end{center}
\end{table}

Equation (\ref{eq:lmi}) can only be solved for a limited selection of
experimental bounds both because of computing time limits, and because
some of the constraints require more refined treatments beside a
simple comparison between theoretical predictions and upper bounds.
In the solution of Eq.~(\ref{eq:lower_bounds}) we have used the 8 DAMA
modulation amplitudes for 1 keVee $\leq E^{\prime} \leq$ 5 keVee~\cite{dama_2018},
and selected energy bins from XENON1T~\cite{xenon_2018}, PICO--60
($C_3F_8$ target) \cite{pico60,pico60_2019}, COSINE--100\cite{cosine_nature},
COUPP~\cite{coupp}, SuperCDMS~\cite{super_cdms_2017} and
PICASSO~\cite{picasso}, as shown in Table~\ref{table:exp_lmi}.

On the other hand, in the exclusion plots in
Figs.~\ref{fig:mchi_sigma_spin_0}, \ref{fig:mchi_sigma_spin_1_2} and
\ref{fig:mchi_sigma_spin_1} we have included an extensive set of
constraints that are representative of the different techniques used
to search for DM: XENON1T~\cite{xenon_2018},
CDMSlite~\cite{cdmslite_2017}, SuperCDMS~\cite{super_cdms_2017},
PICASSO~\cite{picasso}, PICO--60 (using a $CF_3I$ target
~\cite{pico60_2015} and a $C_3F_8$ one \cite{pico60,pico60_2019}), CRESST-II
\cite{cresst_II,cresst_II_ancillary}, DAMA (modulation data
\cite{dama_1998, dama_2008,dama_2010,dama_2018} and average count rate
\cite{damaz}), DarkSide--50 \cite{ds50}.  providing an {\it a
  posteriori} confirmation that the limited set of bounds in
Table~\ref{table:exp_lmi} used to solve Eq.~(\ref{eq:lmi}) did not
miss any relevant constraint.

In the following, if not specified otherwise we adopt
for the energy resolution a Gaussian form, ${\cal
  G}(E^{\prime},E_{ee})=Gauss(E^{\prime}|E_{ee},\sigma)=1/(\sqrt{2\pi}\sigma)exp(-(E^{\prime}-E_{ee})/2\sigma^2)$.
The quenching factor of bolometers (SuperCDMS, CRESST-II) is
assumed to be equal to 1.

\subsection{Xenon: XENON1T}

For XENON1T we have assumed 7 WIMP candidate events in the range of
3PE $ \le S_1 \le $ 70PE, as shown in Fig.~3 of Ref.~\cite{xenon_2018}
for the primary scintillation signal S1 (directly in Photo Electrons,
PE), with an exposure of 278.8 days and a fiducial volume of 1.3 ton
of xenon. 
We have used the efficiency taken from Fig.~1
of~\cite{xenon_2018} and employed a light collection efficiency
$g_1$=0.055; for the light yield $L_y$ we have extracted the best
estimation curve for photon yields $\langle n_{ph} \rangle /E$ from
Fig.~7 in~\cite{xenon_2018_quenching} with an electric field of
$90~{\rm V/cm}$.

For XENON1T experiment we have modeled the energy resolution
combining a Poisson fluctuation of the observed primary signal $S_1$
compared to $<S_1>$ and a Gaussian response of the photomultiplier
with $\sigma_{PMT}=0.5$, so that:

\begin{equation}
{\cal G}_{Xe}(E_R,S)=\sum_{n=1}^{\infty}
Gauss(S|n,\sqrt{n}\sigma_{PMT})Poiss(n,<S(E_R)>),
\label{eq:g_xe}  
\end{equation}

\noindent with $Poiss(n,\lambda)=\lambda^n/n!exp(-\lambda)$.

\subsection{Argon: DarkSide-50}
\label{app:argon}
The analysis of DarkSide-50 \cite{ds50} is based on the ionization
signal extracted from liquid argon with an exposure of 6786.0 kg
days. The measured spectrum for $N_{e^-}<$ 50 (with $N_{e^-}$ the
number of extracted electrons) is shown in Fig. 7 of \cite{ds50}, and
shows an excess for 4 $<N_{e^-}<$7 $N_{e^-}$ compared to a simulation
of the background components from known radioactive contaminants.
Following Ref.\cite{ds50} we have subtracted the background minimizing
the likelihood function: 

\be -2 {\cal L} =\sum_i \frac{(\sigma S_i+\rho
  b_i-x_i)^2}{\sigma_i^2},
\label{eq:bck_chi2}
\ee

\noindent where $i$ represents the energy bin, $x_i$ the measured
spectrum with error $\sigma_i$, while $\sigma S_i$ and $\rho b_i$ are
the DM signal and the background, respectively, with $\sigma$ and
$\rho$ arbitrary normalization factors ($\sigma$ is identified with
the effective WIMP-proton cross section $\sigma_p$). In particular we
obtain the 90\% C.L. upper bound on $\sigma_p$ by taking its profile
likelihood with $-2 {\cal L}- [-2 {\cal L}]_{min}=n^2$ and
$n$=1.28. We take $x_i$, $\sigma_i$ and $b_i$ from Fig.7
of~\cite{ds50}. The ionization yield of argon has been measured only
down to $\lsim$ 10 keVnr, while DS50 uses a model fit to calibration
data. We use the latter as taken from Fig. 6 of \cite{ds50} with a
hard cut at 0.15 keVnr, the lowest energy for which it is provided. We
take the efficiency from Fig. 1 of \cite{ds50}.
\subsection{Germanium: SuperCDMS and CDMSlite}
The latest SuperCDMS analysis \cite{super_cdms_2017} observed 1 event
between 4 and 100 keVnr with an exposure of 1690 kg days. We have
taken the efficiency from Fig.1 of \cite{super_cdms_2017} and the
energy resolution $\sigma=\sqrt{0.293^2+0.056^2 E_{ee}}$ from
\cite{cdms_resolution}. To analyze the observed spectrum we apply the
optimal interval method \cite{yellin}.

For CDMSlite we considered the energy bin of 0.056 keV$<E^{\prime}<$
1.1 keV with a measured count rate of 1.1$\pm$0.2 [keV kg day]$^{-1}$
(Full Run 2 rate, Table II of Ref. \cite{cdmslite_2017}). We have
taken the efficiency from Fig.4 of \cite{cdmslite_2017} and the energy
resolution $\sigma=\sqrt{\sigma_E^2+B E_R+(A E_R)^2}$, with
$\sigma_E$=9.26 eV, $A$=5.68$\times 10^{-3}$ and $B$=0.64 eV from
Section IV.A of~\cite{cdmslite_2017}.
\subsection{Fluorine: COUPP, PICASSO and PICO--60}

Bubble chambers are threshold experiments for which we employ the
nucleation probability:

\begin{equation}
{\cal P}_T(E_R)=1-\exp\left [-\alpha_T\frac{E_R-E_{th}}{E_{th}} \right ].
\label{eq:nucleation_probability}
\end{equation}

COUPP is bubble chamber using a $CF_3I$ target.  For each operating
threshold used in COUPP the corresponding exposure and number of
measured events are summarized in Table \ref{table:coupp}. For
fluorine and carbon we use $\alpha$=0.15 in
Eq.(\ref{eq:nucleation_probability}). For iodine we adopt instead a
step function with nucleation probability equal to 1 above the energy
threshold.

\begin{table}[t]
\begin{center}
{\begin{tabular}{@{}|c|c|c|c|@{}}
\hline
$E_{th}$ (keV) & exposure (kg day) & measured events  \\
\hline
7.8 & 55.8  & 2 \\
11 & 70   & 3 \\
15.5 & 311.7  & 8 \\
\hline
\end{tabular}}
\caption{The operating thresholds with corresponding exposures and
  measured events for COUPP~\cite{coupp}. \label{table:coupp}}
\end{center}
\end{table}

\noindent The PICASSO experiment \cite{picasso} uses $C_4 F_{10}$ as a
target and operated its runs with six energy thresholds. For each
threshold we provide the corresponding number of observed events and
statistical fluctuations in Table \ref{table:picasso} (extracted from
Fig. 4 of Ref.~\cite{picasso}). For the nucleation probability we used
Eq.(\ref{eq:nucleation_probability}) with $\alpha_C$=$\alpha_F$=5.

\begin{table}[t]
\begin{center}
{\begin{tabular}{@{}|c|c|c|c|c|@{}}
\hline
$E_{th}$ (keV) & Event rate (events/kg/day) & Fluctuation \\
\hline
1.0 &   -1.5 & 3.8 \\
1.5 &    -0.2 & 1.0 \\
2.7 &     0.3 & 0.8 \\
6.6 &     -0.8 & 1.8 \\
15.7 &    -1.4 & 2.3 \\
36.8 &     0.3 &1.0 \\
 \hline
\end{tabular}}
\caption{Observed number of events and 1--sigma statistical
  fluctuations (extracted from Fig. 4 of
  Ref. \cite{picasso}) for each operating threshold used in PICASSO.
  \label{table:picasso}}
\end{center}
\end{table}

One of the target materials used by PICO--60 is $C_3F_8$, for which we
used the complete exposure~\cite{pico60_2019} consisting in 1404 kg
day at threshold $E_{th}$=2.45 (with 3 observed candidate events and 1
event from the expected background, implying an upper bound of 6.42
events at 90\%C.L.~\cite{feldman_cousins}) and 1167 kg day keV at
threshold $E_{th}$=3.3 keV (with zero observed candidate events and
negligible expected background, implying a 90\% C.L. upper bound of
2.3 events). For the two runs we have assumed the nucleation
probabilities in Fig. 3 of \cite{pico60_2019}.

\subsection{Fluorine+Iodine: PICO--60}

PICO--60 can also employ a $CF_3I$ target. For the analysis of
Ref.\cite{pico60_2015} we adopt an energy threshold of 13.6 keV and an
exposure of 1335 kg days. The nucleation probabilities for each target
element are taken from Fig.4 in~\cite{pico60_2015}.

\subsection{Sodium Iodide: DAMA and COSINE--100}
For DAMA we consider both the upper bound from the average count rate
(DAMA0) and the latest result for the annual modulation
amplitudes. For DAMA0 we have taken the average count rates from
\cite{damaz} (rebinned from 0.25-keVee- to 0.5-keVee-width bins) from
2 keVee to 8 keVee. We use the DAMA modulation amplitudes normalized
to kg$^{-1}$day$^{-1}$keVee$^{-1}$ in the energy range 1 keVee
$<E^{\prime}<$ 8 keVee from Ref.\cite{dama_2018}. In both cases we
assume a constant quenching factors $q$=0.3 for sodium and $q$=0.09
for iodine, and the energy resolution $\sigma$ = 0.0091
(E$_{ee}$/keVee) + 0.448 $\sqrt{E_{ee}/{\rm keVee}}$ in keV.

The exclusion plot for COSINE--100~\cite{cosine_nature} relies on a
Montecarlo~\cite{cosine_bck} to subtract the different backgrounds of
each of the eight crystals used in the analysis. In
Ref.~\cite{cosine_nature} the amount of residual background after
subtraction is not provided, so we have assumed a constant background
$b$ at low energy (2 keVee$< E_{ee}<$ 8 keVee), and estimated $b$ by
tuning it to reproduce the exclusion plot in Fig.4 of
Ref.~\cite{cosine_nature} for the isoscalar spin-independent elastic
case. The result of our procedure yields $b\simeq$0.13
events/kg/day/keVee, which implies a subtraction of about 95\% of the
background.  We take the energy resolution $\sigma/\mbox{keV}=0.3171
\sqrt{E_{ee}/\mbox{keVee}}+0.008189 E_{ee}/\mbox{keVee}$ averaged over
the COSINE--100 crystals~\cite{cosine_private} and the efficiency for
nuclear recoils from Fig.1 of Ref.~\cite{cosine_nature}. Quenching
factors for sodium and iodine are assumed to be equal to 0.3 and 0.09
respectively, the same values used by DAMA.


\subsection{$CaWO_4$: CRESST-II}
CRESST-II measures heat and scintillation using $CaWO_4$ crystals. We
considered the Lise module analysis from \cite{cresst_II} with energy
resolution $\sigma$=0.062 keV and detector efficiency from Fig. 4
of~\cite{cresst_II_description}. For our analysis we have selected 15
events for 0.3 keVnr$<E_R<$ 0.49 keVnr with an exposure of 52.15 kg
days.


\end{document}